\documentclass[number,sort&compress,3p]{elsarticle}

\usepackage{graphicx}
\usepackage{dcolumn}
\usepackage{bm}
\usepackage{amssymb}
\usepackage{amsmath}
\usepackage{amsfonts}
\usepackage{mathrsfs}
\usepackage{wrapfig}
\usepackage{color}
\usepackage[dvipsnames]{xcolor}
\usepackage{epstopdf}
\usepackage{epsfig}
\usepackage{hyperref}

\pdfoutput=1
\begin{document}
\begin{frontmatter}
\title{Parallel pumping for magnon spintronics: Amplification and manipulation of magnon spin currents on the micron-scale}

\author[SPT,TUKL]{T.~Br\"acher\corref{cor1}}
\ead{thomas.braecher@cea.fr}

\author[TUKL]{P.~Pirro}

\author[TUKL]{B.~Hillebrands}

\cortext[cor1]{Corresponding author}
\address[SPT]{Univ. Grenoble Alpes, CNRS, CEA, INAC-SPINTEC, 17, rue des Martyrs 38054, Grenoble, France}
\address[TUKL]{Fachbereich Physik and Forschungszentrum OPTIMAS, Technische Universit\"at
Kaiserslautern, D-67663 Kaiserslautern, Germany}

\date{\today}

\begin{abstract}
Magnonics and magnon spintronics aim at the utilization of spin waves and magnons, their quanta, for the construction of wave-based logic networks via the generation of pure all-magnon spin currents and their interfacing with electrical charge transport. The promise of efficient parallel data processing and low power consumption renders this field one of the most promising research areas in spintronics. In this context, the process of parallel parametric amplification, i.e., the conversion of microwave photons into magnons at one half of the microwave frequency, has proven to be a versatile tool to excite and to manipulate spin waves. Its beneficial and unique properties such as frequency and mode-selectivity, the possibility to excite spin waves in a wide wavevector range and the creation of phase-correlated wave pairs, have enabled the achievement of important milestones like the magnon Bose Einstein condensation and the cloning and trapping of spin-wave packets. Parallel parametric amplification, which allows for the selective amplification of magnons while conserving their phase is, thus, one of the key methods of spin-wave generation and amplification.

The application of parallel parametric amplification to CMOS-compatible micro- and nanostructures is an important step towards the realization of magnonic networks. This is motivated not only by the fact that amplifiers are an important tool for the construction of any extended logic network but also by the unique properties of parallel parametric amplification. In particular, the creation of phase-correlated wave pairs allows for rewarding alternative logic operations such as a phase-dependent amplification of the incident waves. Recently, the successful application of parallel parametric amplification to metallic microstructures has been reported which constitutes an important milestone for the application of magnonics in practical devices. It has been demonstrated that parametric amplification provides an excellent tool to generate and to amplify spin waves in these systems in a wide wavevector range. In particular, the amplification greatly benefits from the discreteness of the spin-wave spectra since the size of the microstructures is comparable to the spin-wave wavelength. This opens up new, interesting routes of spin-wave amplification and manipulation. In this Review, we will give an overview over the recent developments and achievements in this field.

\end{abstract}
\end{frontmatter}

\tableofcontents

\section{Introduction}
\label{Intro}

The research fields of magnonics and, on a broader scale, magnon spintronics, which includes interfacing to electric charge transport phenomena, aim at the utilization of magnons, the quanta of spin-wave excitation in a magnetic material, as information carriers in a next generation of logic devices beyond conventional CMOS\cite{Magnon-Spintronics,Neusser,Magnon-logical-circuits,Lenk-2011-1,Khitun_2008_1,Magnonics-Krug, Davies-2015-1, TAP-2013,Krawczyk-2014}. This research area has attracted prominent interest in the recent years due to the demonstration of various game-changing devices such as magnonic all-linear time-inverters\cite{Chumak-TI}, magnon transistors\cite{Chumak-MC} and reconfigurable magnonic waveguides\cite{Helmut-2016,Adekunle-2016}, which are just a few, yet instructive, examples. However, for a successful application of magnonics on the micro- and nanometer scale, the research field still faces several challenges. One of the main issues is the spin-wave damping, which is substantial in most conventionally used, CMOS-compatible ferromagnetic metals\cite{Pirro-2011-1, Mockel2011, Mockel2009,Schwarze2013}. While recent studies show the large potential to decrease this damping by material engineering\cite{Sebastian-2012, Pirro-2014-1, Hamadeh-2014-1, Hahn-2014-1}, a reduction of the spin-wave damping comes together with a restriction of the maximum tolerable amplitude of the spin-wave excitation if nonlinear phenomena are to be avoided\cite{Heusler-PRL,SW_basics}. This is, for example, the case if several spin waves are conveying information encoded in their phase in parallel in a magnonic waveguide. Hence, in future magnonic devices, spin waves with small amplitudes and energies in the $\mu\mathrm{eV}$- to $\mathrm{meV}$-range are envisioned to propagate through extended magnonic networks on a chip, conveying information through spin-wave conduits which feature a small, yet finite damping. Therefore, spin-wave amplifiers are of crucial importance to restore the spin-wave amplitude during the course of their propagation as well as to increase the spin-wave intensity to allow for a convenient detection at the output of the network and interconnects to conventional electronics.

In the context of a phase-conserving, mode- and frequency-selective amplification of traveling spin waves, the process of parallel pumping has, so far, proven superior to other approaches of spin-wave amplification, such as the reduction of the spin-wave damping by spin transfer torque (STT)\cite{Mockelwaveguide,Mockeldisc,Gladii-2016-2,An-2014}. Parallel pumping, or parallel parametric amplification, results from the interaction of selected spin waves with a sufficiently large dynamic effective magnetic field, which acts in parallel to the static magnetization and features two times the frequency of these spin waves. Any dynamic effective field modulation with proper frequency can act as source of parametric excitation. Classically, this effective field, referred to as the pumping field, has been provided by a microwave field. In this case, the parametric process creates pairs of magnons by the splitting of microwave photons. Alternatively, the pumping field can be provided by a variation of any magnetic control parameter, such as, e. g., a dynamic variation of the magnetic anisotropy or an incident spin current. One general key feature of the process of parallel pumping is the creation of pairs of magnons which are mutually connected in their phase\cite{SW_basics,LVov:1994ub}. Due to this, the application of parallel pumping goes way beyond a mere amplification of spin waves. Since the relative phase between the created waves can be externally controlled, this allows for a phase-sensitive amplification, which can discriminate logic values encoded in the spin-wave phase\cite{Melkov-Parametric-interaction}. 

In macroscopic systems, the practical application of parallel pumping has suffered from the fact that the preferably amplified mode, referred to as the dominant mode, is generally not the signal carrying fundamental waveguide mode\cite{Rezende-1990,Melkov-Parametric-interaction,Serga:2014hz,Chumak-2009-1,Schaefer-2011}. The latter, which can be conveniently excited externally, features the largest group velocity and net magnetic moment of all modes available in a spin-wave waveguide. Hence, it is the most interesting mode for magnonic applications. However, an efficient amplification of this mode could only be achieved under specific conditions. Recently, it was demonstrated that in microstructured magnonic waveguides and elliptical elements, the dominant mode can be tuned to be the fundamental mode\cite{Braecher-2011-1, Braecher-LocBV, Ulrichs, BobMcM}.  In this Review, we will discuss this phenomenon as well as its consequences on the spin-wave amplification in these microscopic systems.

We briefly review linear spin waves in thin films and microstructured waveguides and introduce the basic equations to describe the effect of parallel parametric amplification on the spin-wave dynamics in Section \ref{Theoback}. Consequently, in Section \ref{PGen}, we present examples of spin-wave generation by parallel parametric amplification of thermal spin waves in several geometries, showing the versatility and the advantages of this technique in comparison to other excitation schemes. In Section \ref{Pamp}, we review the amplification of traveling, externally excited spin-wave packets in magnonic waveguides, addressing the potential and the limitations of this amplification scheme. Finally, we provide an overview over the recent applications and novel concepts of spin-wave manipulation and characterization by means of parallel parametric amplification in Section \ref{Nov}, and, ultimately, a brief conclusion of the Review is provided in Section \ref{Concl}.

\section{Theoretical background}
\label{Theoback}

\subsection{Linear spin waves in microstructures}
\label{SWbasis}

We begin with the presentation of the theory of linear spin waves. It describes the unperturbed eigenstates of the small-amplitude spin-wave excitation in a magnetic material. The spatial confinement in magnetic microstructures leads to the formation of quantized, laterally standing spin-wave modes in these systems. However, already in a thin, laterally extended ferromagnetic film, the finite thickness gives rise to quantized standing spin-wave modes across its thickness, referred to as perpendicular standing spin-wave (PSSW) modes\cite{Kalinikos-1986-Disp,Demo-2009}. In very thin films, experiments involving propagating spin waves and their theoretical description are restricted to the lowest, fundamental thickness mode, which features a quasi-uniform magnetization distribution across the thickness of the film. This is due to the fact that higher modes, which feature a non-uniform magnetization distribution across the thickness, exhibit a large exchange energy and, thus, are found at frequencies beyond the frequency-range typically studied in these experiments. For instance, for a $d = 40\,\mathrm{nm}$ thick film from the magnetic material Ni$_{81}$Fe$_{19}$ (Permalloy) featuring standard parameters\footnote{Saturation magnetization $M_\mathrm{s} = 760\,\mathrm{kA}\,\mathrm{m}^{-1}$, exchange constant $A_\mathrm{ex} = 13\,\mathrm{pJ}\,\mathrm{m}^{-1}$, gyromagnetic ratio $\gamma = 28\,\mathrm{GHz}\,\mathrm{T}^{-1}$.}, the first standing spin wave modes lies at frequencies above $f_1 > 14\,\mathrm{GHz}$ and, thus, more than $10\,\mathrm{GHz}$ above the frequency of the uniform ferromagnetic resonance at $f_0 \approx 3\,\mathrm{GHz}$ for an effective magnetic field of $\mu_0H_\mathrm{eff}=10\,\mathrm{mT}$ within the waveguide. 

\begin{figure}[h]
\center
{\caption{a) Schematic of the lateral amplitude distribution of the first four waveguide modes $n$ = 1 to $n$ = 4 in a spin-wave waveguide. b) Spin-wave dispersion relation for an effectively $2.3\,\mu\mathrm{m}$ wide, $40\,\mathrm{nm}$ thick Ni$_{81}$Fe$_{19}$ spin-wave waveguide in the Damon-Eshbach geometry ($\mathbf{M}\perp\mathbf{k}_{||})$. c) Spin-wave dispersion relation for an effectively $2.7\,\mu\mathrm{m}$ wide, $40\,\mathrm{nm}$ thick Ni$_{81}$Fe$_{19}$ spin-wave waveguide in the backward-volume geometry ($\mathbf{M}||\mathbf{k}_{||})$. The dashed lines represent the exchange-dominated spin waves. For both geometries, $\mu_0H_\mathrm{eff}=25\,\mathrm{mT}$ has been assumed. Please note the different scales in $k_\mathrm{||}$ in b) and c).} 
 \label{Fig:Theo-SW1}}
{\includegraphics[width=1.0\textwidth]{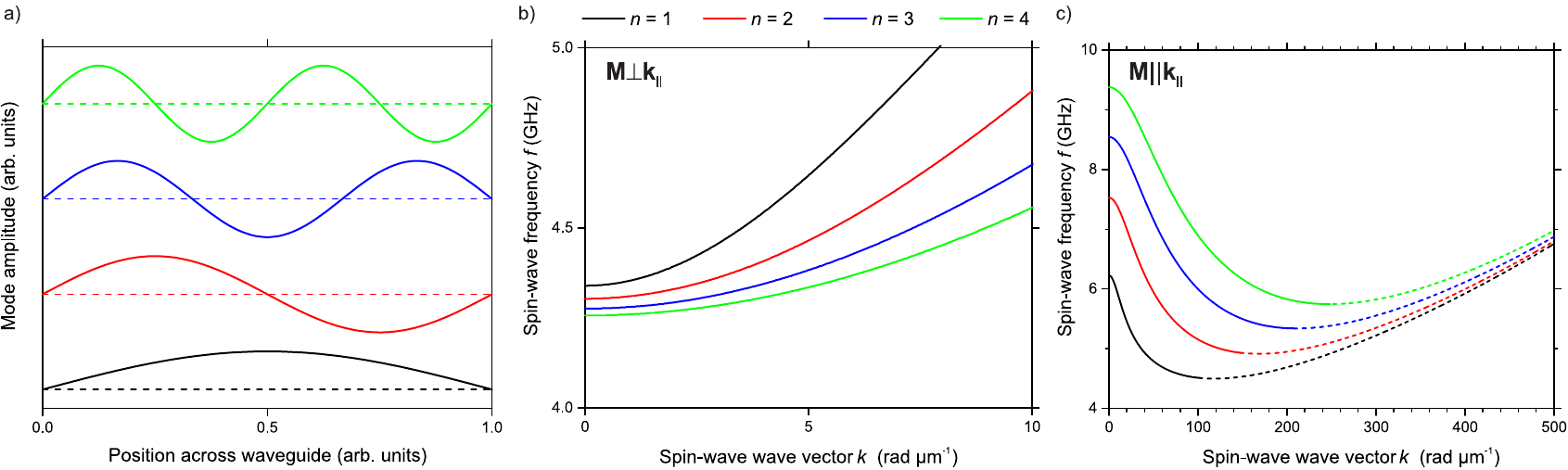}}
\end{figure}

For the uniform mode in an isotropic, in-plane magnetized film, the spin-wave dispersion relation for spin waves with $|\mathbf{k}| \cdot d<1$ is of the compact form\cite{Kalinikos-1986-Disp,Krivosik:2010tz}
\begin{align}
\label{Eq:Disp-tf}
\omega (\mathbf{k}) &= \sqrt{(\omega_\mathrm{H} + \omega_\mathrm{M}\lambda_\mathrm{ex}k^2)\left(\omega_\mathrm{H} + \omega_\mathrm{M}\lambda_\mathrm{ex}k^2+\omega_\mathrm{M}\left(1+g_k(\mathrm{sin}^2(\theta_k)-1)+\frac{\omega_\mathrm{M} g_k(1-g_k)\mathrm{sin}^2(\theta_k)}{\omega_\mathrm{H}+\omega_\mathrm{M}\lambda_\mathrm{ex}k^2}\right)\right)}.
\end{align}
Here, $k = |\mathbf{k}|$ is the magnitude of the spin-wave wavevector $\mathbf{k}$ in the film plane, $\omega_\mathrm{H} = \gamma \mu_0 H_\mathrm{eff}$ with the effective field $\mu_0 H_\mathrm{eff}$ inside the waveguide, $\omega_\mathrm{M} = \gamma \mu_0 M_\mathrm{s}$ with the saturation magnetization $M_\mathrm{s}$, $\lambda_\mathrm{ex}=2A_{ex}/(\mu_0M_\mathrm{s}^2)$ with the exchange constant $A_{ex}$, $\theta_k$ denotes the angle between the spin-wave wavevector $\mathbf{k}$ and the magnetization direction $\mathbf{M}$ and $g_k = 1-(1-\exp(-kd))/(kd)$. If the thin film is structured into a spin-wave waveguide, i.e., a thin rectangular slab, an additional quantization of the wavevector $\mathbf{k}$ is introduced across the width of the waveguide. This gives rise to the formation of waveguide modes. In the geometries reported in this Review, these modes are described reasonably well by a continuous wavevector $k_{||}$ along the waveguide long axis and an approximately cosinusoidal profile across the width of the waveguide. The latter is sketched in Fig.~\ref{Fig:Theo-SW1}, where the amplitude distribution of the first four waveguide modes is schematically shown. 

An approximate derivation of the spin-wave dispersion of the waveguide modes can be obtained by using Eq.~\ref{Eq:Disp-tf} with a fixed wavevector component $k_\perp = n\cdot\pi/w_\mathrm{eff}$ along the short axis of the waveguide, which results in a continuous variation of the angle $\theta_k$ as the wavevector component $k_{||}$ along the propagation direction is changed. Here, $n$ denotes the mode number of the waveguide modes and $w_\mathrm{eff}$ the effective width of the waveguide. This effective width can deviate from the geometrical width of the waveguide due to the boundary conditions of the static and the dynamic magnetization at the waveguide edges\cite{Guslienko-eff-width,Joseph-1965,Bayer-2006}. This approach has been successfully applied in, e.g., Refs. \cite{Vogt-PR, Pirro-2011-1,Bayer-2006} to obtain the approximate spin-wave frequencies. In the two commonly studied geometries, namely in the transversely magnetized waveguide $\mathbf{k}_{||} \perp \mathbf{M}$ (often also referred to as \textit{Damon-Eshbach}-geometry) and the longitudinally magnetized waveguide $\mathbf{k}_{||} || \mathbf{M}$ (\textit{backward-volume}-geometry), the angle $\theta_k$ and the resulting dispersion relations are then calculated using $\theta_k^{\mathrm{DE}}(k_{||}) = \mathrm{arctan}({k_\perp}/{k_{||}})$ and $\theta_k^{\mathrm{BV}}(k_{||}) = \mathrm{arctan}({k_{||}}/{k_\perp})$, respectively. Figure \ref{Fig:Theo-SW1}~(b) shows the dispersion of the first four waveguide modes $n=1$ to $n=4$  in a $w_\mathrm{eff} = 2.3\,\mu\mathrm{m}$ wide, $d = 40\,\mathrm{nm}$ thick transversely magnetized Ni$_{81}$Fe$_{19}$ waveguide for an effective field of $\mu_0H_\mathrm{eff}=25\,\mathrm{mT}$ in the waveguide. Figure \ref{Fig:Theo-SW1}~(c) shows the resulting dispersion in a $w_\mathrm{eff} = 2.7\,\mu\mathrm{m}$ wide, $d = 40\,\mathrm{nm}$ thick longitudinally magnetized waveguide. Both effective widths correspond to a geometrical width of $w = 2.5\,\mu\mathrm{m}$ and the same material parameters of Ni$_{81}$Fe$_{19}$ as stated above have been used. 

\subsection{Parallel pumping}
\label{Theopump}

\subsubsection{Basic equations - $S$-theory}
\label{Stheo}

\begin{figure}[h]
\center
{\caption{a) Schematic of the elliptical magnetization trajectory in a thin film. This trajectory gives rise to the longitudinal dynamic magnetization component $m_l$ which interacts with the microwave pumping field $h_{2f}$. b) The (adiabatic) parallel pumping process in the particle picture: A photon with negligible wavevector splits into a pair of counter-propagating magnons, leading to the formation of the signal and idler wave at one-half of the photon frequency. The solid dark blue line represents the spin-wave dispersion of the fundamental waveguide mode.} 
 \label{Fig:Theo-Pamp1}}
{\includegraphics[width=0.8\textwidth]{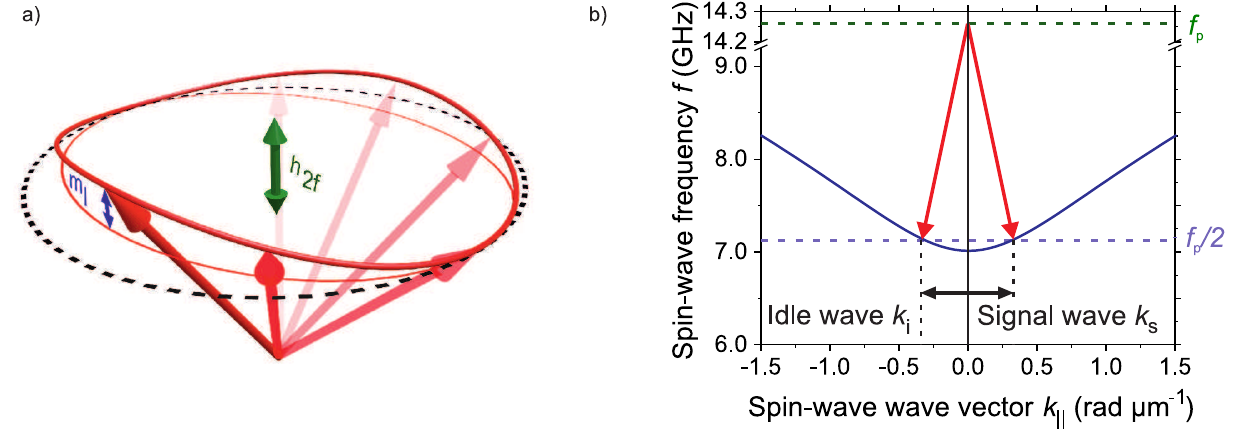}}
\end{figure}

The process of parallel parametric amplification or parallel pumping can be realized by the interaction of a microwave pumping field $\mu_0h_\mathrm{p}$, which is applied along the static magnetization, with the longitudinal component of the dynamic magnetization in a magnetic material. In an in-plane magnetized magnetic thin film, this dynamic longitudinal component is non-zero as a consequence of the shape anisotropy, which results in an elliptic magnetization trajectory, and the fact that the length of the magnetization vector is preserved on the timescale of the magnetization precession\cite{SW_basics} (cf. Fig.~\ref{Fig:Theo-Pamp1}~(a)). In the particle picture, the process of parallel pumping is described by the splitting of microwave photons with frequency $f_\mathrm{p} = f_\mathrm{sw1}+f_\mathrm{sw2}$ into pairs of magnons with frequency $f_\mathrm{sw1}$ and $f_\mathrm{sw2}$\cite{Schloemann-1960-1} under frequency and momentum conservation (cf. Fig.~\ref{Fig:Theo-Pamp1}~(b)). Hereby, the decaying microwave photons provided by the pumping field constitute an energy flow into the magnon system which counteracts the magnon damping. If the inserted energy per unit time exceeds the loss rate $\omega_\mathrm{r}$ of the pumped magnon group, the intensity of the magnon group rises exponentially in time. This compensation is referred to as the parametric instability, which, consequently, is a threshold process which requires a minimum threshold field $\mu_0h_\mathrm{p,th}$. In the following, a short summary of the main findings of the basic $S$-theory is given\cite{LVov:1994ub}. This theory provides a description of the thresholds of the parametric instability as well as of the temporal evolution of the spin-wave intensity when the pumping field is applied. Moreover, it has successfully provided a description of the steady state reached in macroscopic systems under parallel pumping if the applied pumping field $\mu_0h_\mathrm{p}$ does not exceed the threshold field $\mu_0h_\mathrm{p,th}$ significantly.

Assuming a simple harmonic time dependence of the pumping field $\mu_0h_\mathrm{p}(t) = \mu_0h_\mathrm{p} \mathrm{cos}(\omega_\mathrm{p} t)$ with frequency $\omega_\mathrm{p} = 2 \pi f_\mathrm{p}$ without a spatial confinement, and assuming that this field is coaligned with the static magnetization and an external bias field $H_\mathrm{ext}$~(i.e., $\mu_0\mathbf{H_\mathrm{ext}} = \mu_0(H_\mathrm{ext} + h_\mathrm{p})\mathbf{\hat{e}_z})$, the Hamiltonian associated with the pumping process is given by \cite{LVov:1994ub,Krivosik:2010tz}\footnote{For simplicity, it is assumed that no other effective field components occur and, thus, $\mathbf{H}_\mathrm{eff} = \mathbf{H}_\mathrm{ext}$.}.:
\begin{align}
\label{Eq:U2-PP}
\mathscr{H}^{(2),\mathrm{PP}} &= \frac{1}{2}\sum_\mathbf{k}{\mu_0h_\mathrm{p} V(\mathbf{k}) c^*_\mathbf{k} c^*_{-\mathbf{k}} \mathrm{e}^{-\omega_\mathrm{p} t}+ \mathrm{c.c.}},
\end{align}
with the coupling parameter $V(\mathbf{k})$, which describes the coupling of the pumping field $\mu_0h_\mathrm{p}$ to the spin-wave mode $c_\mathbf{k}$. The sum is taken over all modes $\mathbf{k}$ interacting with the pumping field. The fact that the pumping mainly interacts with wave pairs with $\pm \mathbf{k}$ is a consequence of linear momentum conservation:
\begin{align}
\label{Eq:Momcons-adia}
\mathbf{k}_1 + \mathbf{k}_2 = \mathbf{k}_\mathrm{p} \approx 0.
\end{align} 
Here, $\mathbf{k}_\mathrm{p}$ denotes the wavevector of the pumping field, which, in the absence of a strong spatial confinement of the pumping field, is approximately zero due to the long wavelength of the microwave. 
In addition to the interaction of the magnons with the pumping field, the interaction of the created magnons with each other has a large influence on the spin-wave dynamics under parallel pumping. This interaction is described by the four-wave interaction between the generated spin-wave pairs. In many cases, this interaction provides the main mechanism for a saturation of the amplification. In general, it is described by the Hamiltonian
\begin{align}
\label{Eq:U4}
\mathscr{H}^{(4)} = \sum_{\mathbf{k}_1+\mathbf{k}_2 = \mathbf{k}_3 + \mathbf{k}_4}T(\mathbf{k}_1, \mathbf{k}_2, \mathbf{k}_3, \mathbf{k}_4)c^*_{\mathbf{k}_1}c^*_{\mathbf{k}_2}c_{\mathbf{k}_3}c_{\mathbf{k}_4},
\end{align}
where the matrix element $T(\mathbf{k}_1, \mathbf{k}_2, \mathbf{k}_3, \mathbf{k}_4)$ describes the four wave interaction between a pair of waves with $\mathbf{k}_1$ and $\mathbf{k}_2$ with a pair of waves featuring $\mathbf{k}_3$ and $\mathbf{k}_4$. By solving the Hamilton equations including the Hamiltonian $\mathscr{H}^{(\mathrm{pp})}= \mathscr{H}^{(2),\mathrm{pp}}+\mathscr{H}^{(4)}$, which describes the spin-wave interaction with the pumping field, together with the common Hamiltonian $\mathscr{H}^{(2)}$ describing the linear spin-wave dynamics, one obtains the basic equations of the $S$-theory\cite{LVov:1994ub}:
\begin{align}
\label{Eq:S-n}
\frac{\partial n(\mathbf{k},t)}{\partial t}& = 2n(\mathbf{k},t)\left[-\omega_\mathrm{r}(\mathbf{k})+\mathrm{Im}\left\{P^*(\mathbf{k},t)\mathrm{e}^{\mathrm{i}\psi(\mathbf{k},t)}\right\}\right],\\
\frac{\partial \psi(\mathbf{k},t)}{\partial t}& = \omega_\mathrm{NL}(\mathbf{k})-\frac{\omega_\mathrm{p}}{2}+\mathrm{Re}\left\{P^*(\mathbf{k},t)\mathrm{e}^{\mathrm{i}\psi(\mathbf{k},t)}\right\},\label{Eq:S-psi} \\
P(\mathbf{k},t)& = \mu_0h_\mathrm{p} V(\mathbf{k})+\sum_{\mathbf{k}_1}S(\mathbf{k},\mathbf{k}_1)n(\mathbf{k}_1,t)\mathrm{e}^{-\mathrm{i}\psi(\mathbf{k},t)},\label{Eq:S-P} \\
\omega_\mathrm{NL}& = \omega(\mathbf{k})+\sum_{\mathbf{k}_1}T(\mathbf{k},\mathbf{k}_1)n(\mathbf{k}_1,t).\label{Eq:S-NL}
\end{align}
Here, $n(\mathbf{k},t)=c^*_\mathbf{k} c_\mathbf{k}$ denotes the density of the magnon group $c_\mathbf{k}$, $\psi(\mathbf{k},t)$ represents the phase of the wave pairs with $\pm \mathbf{k}$, i.e. $\psi(\mathbf{k},t) = \phi(\mathbf{k},t)+\phi(-\mathbf{k},t)$, $P$ is the complete pumping. The latter is constituted of the energy flux by the external pumping $\mu_0h_\mathrm{p} V(\mathbf{k})$ and the energy flux by the \textit{internal pumping}, which describes the energy exchange between the wave pairs, and which is given by the sum in Eq.~\ref{Eq:S-P}. 
The fourth equation describes the nonlinear frequency $\omega_\mathrm{NL}$, which can deviate from the linear frequency $\omega(\mathbf{k})$ of the created spin waves due to the creation of spin-wave pairs during the pumping. In the following, this shift is neglected since, in the discussed experiments in microstructures, only a low supercriticality $h_\mathrm{p}/h_\mathrm{p,th}$ is treated and, consequently, the nonlinear frequency shift is low. The summation is performed over all wave-pairs interacting with the pumping field. 
The parameter $S(\mathbf{k},\mathbf{k}_1)$, which is the name-giving parameter of the $S$-theory, describes the four-magnon coupling between waves with $\pm \mathbf{k}$, i.e., $S(\mathbf{k},\mathbf{k}_1)=S^*(\mathbf{k},\mathbf{k}_1)=T(\mathbf{k},\mathbf{k}_1,-\mathbf{k},-\mathbf{k}_1)/2$. It should be noted that the phase $\psi(\mathbf{k},t)$ of the wave pairs is well defined by Eq.~\ref{Eq:S-psi}, while the individual phases $\phi(\mathbf{k},t)$ and $\phi(-\mathbf{k},t)$ can be selected spontaneously from the thermal spin-wave background or predefined by an external excitation. 

\begin{figure}[h]
\center
{\caption{a) Calculated threshold fields for the first four waveguide modes in a transversely magnetized Ni$_{81}$Fe$_{19}$ waveguide (dimensions see text). The inset shows the threshold fields at low frequencies in an enlarged scale. b) Calculated threshold fields for the first four waveguide modes in a longitudinally magnetized Ni$_{81}$Fe$_{19}$ waveguide (dimensions see text). The dashed lines mark the exchange spin-wave branches.} 
 \label{Fig:Theo-Pamp2}}
{\includegraphics[width=0.8\textwidth]{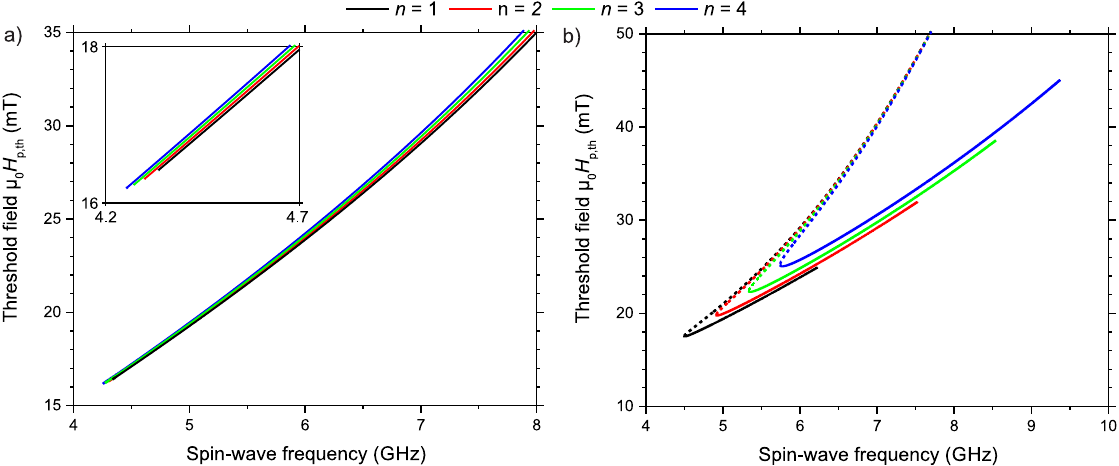}}
\end{figure}

Initially, i.e., before the parametric amplification starts, the system can be treated as linear and the four-wave interaction can be neglected, i.e., $S(\mathbf{k},\mathbf{k}_1)=0$. Under these conditions, Eq.~\ref{Eq:S-n} reduces to a simple rate equation, where the pumping is acting as a source of magnons which are dissipated due to their interaction with the electron and phonon system. From this equation the threshold condition can be obtained\cite{LVov:1994ub}:
\begin{align}
\label{Eq:Thresh1}
\nu_\mathbf{k} = -\omega_\mathrm{r} (\mathbf{k})+ \sqrt{|\mu_0h_\mathrm{p} V(\mathbf{k})|^2 - \left(\omega(\mathbf{k}) - \frac{\omega_\mathrm{p}}{2}\right)^2} \overset{!}{=} 0.
\end{align}
Here, $\nu_\mathbf{k}$ accounts for the incremental change of the magnon density $n(\mathbf{k},t) = c_\mathbf{k}^*(t)c_{\mathbf{k}}(t)$ in the form of
\begin{align}
\label{Eq:Intvicth}
 n(\mathbf{k},t) = n_0(\mathbf{k}) \mathrm{e}^{(2\nu_\mathbf{k} t)},
\end{align} 
which holds on short timescales for powers in the vicinity of the threshold (i.e. for small $\nu$). $n_0 (\mathbf{k})$ denotes the starting density of the waves. Thus, the threshold field $\mu_0h_\mathrm{p,th}$ is given by
\begin{align}
\label{Eq:hcritf}
\mu_0h_\mathrm{p,th} (\mathbf{k}) = \sqrt{\frac{\omega_\mathrm{r}(\mathbf{k})^2 + \left(\omega(\mathbf{k}) - \frac{\omega_\mathrm{p}}{2}\right)^2}{|V(\mathbf{k})|^2}}.
\end{align}
In general, the last term in the numerator becomes minimum for $\omega(\mathbf{k}) = \omega_\mathrm{p}/2$ and, consequently, the pumping is degenerate, unless large variations of $V(\mathbf{k})$ or $\omega_\mathrm{r}(\mathbf{k})$ around $\omega(\mathbf{k}) = \omega_\mathrm{p}/2$ make a non-degenerate splitting favorable. Thus, in the general case, the magnon pairs with the minimal threshold feature half the frequency of the pumping field\footnote{Otherwise, Eq.~\ref{Eq:hcritf} reads 
\begin{align}
\mu_0h_\mathrm{p,th} (\mathbf{k}_1,\mathbf{k}_2) = \sqrt{\frac{\omega_\mathrm{r}(\mathbf{k}_1)\omega_\mathrm{r}(\mathbf{k}_2) + \left(\omega(\mathbf{k}_1) - \frac{\omega_\mathrm{p}}{2}\right)\left(\omega(\mathbf{k}_2) - \frac{\omega_\mathrm{p}}{2}\right)}{|V(\mathbf{k}_1)||V(\mathbf{k}_2)|}},
\end{align}
and similar changes also apply to the other equations stated above.}. 
In this case, the threshold field is simply given by
\begin{align}
\label{Eq:Threshold}
\mu_0h_\mathrm{p}(\mathbf{k}) = \frac{\omega_\mathrm{r}(\mathbf{k})}{|V(\mathbf{k})|}.
\end{align}
Therefore, the mode with the lowest relaxation frequency $\omega_\mathrm{r}(\mathbf{k})$ and the highest coupling $V(\mathbf{k})$ features the lowest threshold. This mode is referred to as the critical or the dominant mode. From the spin-wave dispersion, the relaxation frequency $\omega_\mathrm{r}$ caused by the Gilbert damping, which is dominating in the case of metallic microstructures, can be derived via\cite{Stancil}
\begin{align}
\label{Eq:TauComp}
\omega_\mathrm{r} &= \alpha \left(\omega_\mathrm{H} + \omega_\mathrm{M}\lambda_\mathrm{ex}k^2 + \frac{\omega_\mathrm{M}}{2}(1+g(\mathbf{k})(\sin^2(\theta(\mathbf{k}))-1))\right),
\end{align}
and the coupling parameter $V$ follows to\cite{SW_basics}
\begin{align}
\label{Eq:V}
V(\mathbf{k}) = \frac{\gamma\omega_\mathrm{M}}{4 \omega(\mathbf{k})}[g_k \sin^2(\theta_k)-(1-g_k)],
\end{align}
which is basically a measure for the ellipticity of the magnetization trajectory. Since the ellipticity is maximal for the uniform precession and decreases with increasing wavevector, the coupling decreases as the mode-number and the wavevector component in the propagation direction increase. The resulting threshold fields using the parameters stated above for the calculation of the spin-wave dispersion relations and a Gilbert damping parameter of $\alpha = 0.008$ are presented in Figs.~\ref{Fig:Theo-Pamp2}~(a) and \ref{Fig:Theo-Pamp2}~(b) for the first four waveguide modes of a transversely and a longitudinally magnetized Ni$_{81}$Fe$_{19}$ waveguide, respectively. As can be seen from the panels, the threshold fields are fairly similar in magnitude and, generally, the lowest available waveguide mode $n$ features the lowest threshold field (cf. inset in Fig.~\ref{Fig:Theo-Pamp2}~(a) for the transversely magnetized waveguide). Hence, it is expected that the fundamental waveguide mode $n = 1$ is the dominant mode in these scenarios. This is a unique feature of microstructured elements and a major difference to the previously studied macroscopic samples (see, e.g., Refs. \cite{Rezende-1990,Melkov-Parametric-interaction,Serga:2014hz,Chumak-2009-1,Schaefer-2011}). Since this mode is the most interesting mode for magnonics as it features the largest net magnetic moment and the largest group velocity among all waveguide modes (compare slopes of dispersion in Fig.~\ref{Fig:Theo-SW1}), this renders parallel pumping particularly interesting in microstructured spin-wave conduits.

An alternative way to look at the threshold is the energy balance of the spin-wave system. For this, one should compare the energy flux $W^+$ into the spin-wave system by the external pumping field with the energy flux $W^-$, which describes the losses of this spin-wave mode per unit-time. Initially, i.e., before wave pairs are generated, for a pair $\pm \mathbf{k}$ with the assumptions $\omega(\mathbf{k}) = \omega(\mathbf{-k})$ and $\omega_\mathrm{p} = 2\omega(\mathbf{k})$, these are given by\cite{LVov:1994ub}:
\begin{align}
\label{Eq:W+}
W^+& = -\frac{\partial \mathscr{H}}{\partial t} = 2\omega_\mathrm{p}|\mu_0h_\mathrm{p} V(\mathbf{k})|n(\mathbf{k},t)\mathrm{sin}(\psi_\mathrm{pump}(t)-\psi(\mathbf{k},t)),\\
W^-& = 2 \omega_\mathrm{r}(\mathbf{k}) \omega_\mathrm{p} n(\mathbf{k},t) 
\label{Eq:W-}
\end{align}
with the pumping phase $\psi_\mathrm{pump}(\mathbf{k},t)$, which is given by $\psi_\mathrm{pump}(\mathbf{k},t) = \arg(\mu_0h_\mathrm{p} (t) V(\mathbf{k}))$ at the threshold. Again, $\psi(\mathbf{k},t) = \phi(\mathbf{k},t)+\phi(-\mathbf{k},t)$ denotes the phase of the spin-wave pair. From this, it follows directly\footnote{Since, in the absence of a strong nonlinearity, $\omega_\mathrm{p} = 2\omega(\mathbf{k})$ there is no temporal dynamic of the phase difference.} that the energy flux into the mode $n(\mathbf{k},t)$ is maximal if 
\begin{align}
\label{Eq:Phaserel}
\psi(\mathbf{k}) = \phi(\mathbf{k},t)+\phi(-\mathbf{k},t) = \psi_\mathrm{pump} + \pi/2, 
\end{align}
i.e., the phase between the wave pair and the pumping field is shifted by $\pi/2$. Under this condition, the threshold, which is defined by $W^+ = W^-$, is reached if:
\begin{align}
\label{Eq:Threshold-gen}
|\mu_0h_\mathrm{p} V(\mathbf{k})| = \omega_\mathrm{r}(\mathbf{k}),
\end{align}
which describes an equivalent threshold condition to Eq.~\ref{Eq:Threshold}. Another very important feature can be seen from Eq.~\ref{Eq:W+}: The energy flux into the mode $\mathbf{k}$ is proportional to its density $n(\mathbf{k})$. Thus, the mode $\mathbf{k}$ needs to be initially populated in order to be amplified. At room temperature, this initial population is generally provided by a homogeneous thermal excitation of practically all spin-wave modes $\mathbf{k}$ in the investigated frequency range, since $\hbar \omega(\mathbf{k}) \ll k_\mathrm{B} T$. Alternatively, a (in general significantly larger) population of spin waves at a desired frequency can be realized by an external excitation with driving frequency $\omega$.

As depicted in Fig.~\ref{Fig:Theo-Pamp1}~(b), due to the negligible linear momentum of the microwave photons $k_\mathrm{p} = 0$, parallel pumping a priori leads to the formation of counter-propagating waves with $\pm k$ in order to fulfill momentum conservation. In the context of parametric amplification of externally excited, signal carrying waves, these two waves are commonly referred to as the signal ($k_\mathrm{s}$) and idler ($k_\mathrm{i} = -k_\mathrm{s}$) wave, which share a mutual phase relation according to Eq.~\ref{Eq:Phaserel}. Hereby, the signal wave is the wave running in parallel to the input waves whereas the idler wave runs in the opposite direction. This process is commonly referred to as the adiabatic regime of parallel pumping. However, by providing a spatial confinement to the pumping field, it is possible to obtain a pumping source with an effective wavevector spectrum $k_\mathrm{p} \gg 0$. In this so called nonadiabatic regime of parallel pumping\cite{Melkov-nonadiabatic,Melkov-Parametric-interaction,Serga_Phase-controll}, the finite momentum provided by the localization allows for a splitting of microwave photons with $k_\mathrm{p} >0$ into pairs of co-propagating waves following
\begin{align}
\label{Eq:conservation-nonadiabatic}
{k_\mathrm{p}=k_\mathrm{s}+k_\mathrm{i}=2k_\mathrm{s}~~\rightarrow~~k_\mathrm{i}=k_\mathrm{s}}.
\end{align}

\subsubsection{Limiting mechanisms beyond the parametric instability threshold}
\label{Theo-PPA-Sat}

Several mechanisms can limit the amplitude of the amplified spin waves. The most important ones are the dephasing mechanism which has been originally introduced by Zhakarov, L'Vov and Starobinets and the nonlinear damping of the generated waves. In particular, the dephasing mechanism has proven to be responsible for most observations made in the material yttrium iron garnet (YIG), where the phenomenon of parallel pumping was extensively studied in bulk and thin-film samples (see, e.g., Refs. \cite{SW_basics,Serga:2014hz,2011PhRvL.106u6601S,Vitaliy,YIGmagnonics}). The dephasing mechanism is already incorporated into Eqs.~\ref{Eq:S-n}-\ref{Eq:S-NL} and it is based on the fact that the generated waves provide a mutual pumping field by themselves, named the internal pumping\footnote{An additional nonlinear frequency shift due to an extensive magnon generation also adds to the dephasing mechanism since $\omega_\mathrm{NL}$ can deviate from $\omega_\mathrm{p}/2$ (cf. Eq.~\ref{Eq:S-psi}).}. In a simple picture, if a certain population $n(\mathbf{k})$ is already established, a new magnon pair $\pm \mathbf{k}$ which is supposed to be generated feels the influence of the external pumping field as well as the one of the internal pumping field, which is proportional to $n(\mathbf{k},t)$. Due to the fact that the phase of the already generated wave-pairs differs from the phase of the external pumping initially by $\pi/2$, the newly generated magnons now have to adjust their phase to the phase of the effective pumping field. By doing so, the energy flux $W^+$ from the pumping field into the magnon group decreases until, at some point, at a pumping field above the threshold field $\mu_0h_\mathrm{p,th}$, $W^+$ matches the flux $W^-$ and the spin-wave density $n(\mathbf{k})$ saturates. 

A numerical demonstration of the influence of this process under the assumption of the presence of two modes $\mathbf{k}_1$ and $\mathbf{k}_2$ can, for example, be found in Refs. \cite{Chumak-2009-1,PhD-Schaefer}. In particular, these works address the coexistence of two modes in the system of which one is the signal mode $n_\mathrm{Sig}$ and the other one is the dominant mode $n_\mathrm{Dom}$. Here, the dominant mode is supposed to be the thermally excited exchange wave featuring the lowest instability threshold while the signal mode is excited initially at a much larger level, i.e., $n_\mathrm{Sig} \gg n_\mathrm{Dom}$. In return, it features a larger damping $\omega_\mathrm{r,\mathrm{Sig}}>\omega_\mathrm{r,\mathrm{Dom}}$. If now a pumping field, which is sufficiently large to exceed the threshold of the signal mode, is applied, the signal group is initially favored due to its higher starting population. However, after a certain period, the dominant mode will end up with a larger population due to its larger intensity increment $\nu(\omega_\mathrm{r})$ per unit time (cf. Eq.~\ref{Eq:Thresh1}). Once the amplitudes of the two waves have increased beyond a certain critical level, the dephasing mechanism limits the magnon density in the steady state. In particular, in the steady state the internal pumping field has decreased the total pumping field to the point where $W^+_\mathrm{Dom} = W^-_\mathrm{Dom}<W^-_\mathrm{Sig}$, where the last inequality is based on the fact\footnote{Similar arguments hold for the coupling parameter - by definition, $W^+$ of the dominant group is largest and/or $W^-$ is smallest, since it features the lowest $\omega_\mathrm{r}$ and/or the largest $V$.} that $\omega_\mathrm{r,\mathrm{Sig}}>\omega_\mathrm{r,\mathrm{Dom}}$. Thus, the losses of the signal group are smaller than the inserted energy and it will start to decay in time. Hence, on the long run, only the dominant mode survives in the steady state, which is a general feature of this mode that also holds if more than two modes are investigated. This concept puts important restrictions on the amplification of traveling spin waves by the mechanism of parametric amplification, especially if the signal mode is not the dominant mode.

The second important limiting mechanism of the spin-wave amplitude is the nonlinear damping of the spin waves themselves, i.e., a change of the spin-wave relaxation frequency $\omega_\mathrm{r}$ due to magnon-magnon interaction. This can be included by an additional term $\omega_\mathrm{r}' \propto -n(\mathbf{k})$ in the rate equations\cite{LVov:1994ub}. While in most experiments performed in bulk materials and thin films this is only a side effect which appears at a rather large supercriticality $h_\mathrm{p} \gg h_\mathrm{p,th}$, the nonlinear damping can play a more decisive role in microstructures such as transversely magnetized waveguides. In particular, this can be the case because, depending on the geometry, the waveguide modes in these waveguides can exhibit very similar relaxation frequencies $\omega_\mathrm{r}$ and coupling parameters $V(\mathbf{k})$ and, thus, feature a very similar threshold (cf. Fig.~\ref{Fig:Theo-Pamp2}). Hence, the usual assumption that only one spin-wave mode is excited by the parametric amplification is not valid anymore. Even at very low supercriticalities a whole set of waveguide modes is potentially amplified. Before the aforementioned saturation due to the phase-mechanism sets in, where only the dominant mode is maintained by the action of the pumping field, this increases the amplitude of all of these modes. For a four-magnon process, the coupling between the involved modes is proportional to the product of their amplitudes (cf., for example, \cite{PhD-Pirro}) and the threshold for such processes is inversely proportional to the spin-wave damping. Thus, the fact that the damping of several modes $n_1$, $n_2$, ... is compensated by the pumping field also increases the nonlinear damping of each of these modes due to the four-magnon interaction with all other modes which are excited in this transition phase. Consequently, it is expected that nonlinear damping can also have a significant influence on the amplitude limitation in microstructures.

It should be noted that both of these limitation mechanisms imply that the intensity level at which the parametrically amplified spin waves saturate depends on the supercriticality. The higher the applied pumping power above the threshold, the higher the saturation level which can be maintained in the steady state under the act of parallel pumping. Thus, to achieve a large spin-wave density of the amplified waves, a large supercriticality is generally favorable.

\subsubsection{Limitation of the applicability of the $S$-theory for the investigated microstructures}
\label{Theo-PPA-Lim}

As mentioned above, the fact that in the investigated waveguides generally more than one mode is excited for very small supercriticalities already limits the simple application of the $S$-theory described above. Moreover, the $S$-theory has been derived explicitly for spatially homogeneous samples and pumping fields. Therefore, it cannot take into account the finite sizes of the analyzed waveguides or the resulting lateral quantization effects. The prediction of the thresholds, however, is expected to be correct since it comes from the very general energy balance which can - at least via the spin-wave dispersion - account for the effects arising from the quantization. In fact, it has been demonstrated in \cite{Heusler-PRL} and \cite{PhD-Pirro} that the general Hamilton approach sketched above can describe the thresholds and the critical modes for spin-wave instabilities in waveguides reasonably well. Since the treatment of the parallel-pumping instability is very similar to the treatment of these spin-wave interactions, one can expect that the prediction of the thresholds is valid. Still, the predictions of the $S$-theory for the steady state and the limitation mechanisms have to be considered rather as a tool for a qualitative understanding of the occurring phenomena. 

An additional side effect of the quantization should be noted at this point: Only a limited number of modes exists at a given frequency $\omega(\mathbf{k})$ as a consequence of the lateral quantization and due to the low film thicknesses. 
Thus, it is possible to realize the case in which the fundamental waveguide mode constitutes the dominant mode as well as the signal mode. As mentioned above, this is hardly ever the case in macroscopic samples. In these samples typically some exchange mode features the lowest threshold\cite{PhD-Schaefer,Melkov-Parametric-interaction}. Thus, after a certain time, this mode diminishes the intensity of the signal wave and suppresses the information stored in this mode. In microstructures, this implies that, at least according to the phase mechanism, the steady-state can be reached for the signal waves and the information carried by these waves can, this way, be stored in the spin-wave system by the pumping field.

\section{Parametric generation - amplification of thermal waves}
\label{PGen}

Parametric generation is referred to as the amplification of thermal spin waves\cite{2011PhRvL.106u6601S,Serga:2014hz,Serga-2003-PP,Kostylev-1995-1,Vendik-1977-PP,Kalinikos-1989-PP,Kalinikos-1985-PP,Kalinikos-1985-PP2,Wiese-1994}. Spin waves are always thermally excited at room temperature due to their $\mu\mathrm{eV}$ excitation energy. The pumping field couples to the longitudinal dynamic component provided by these thermal waves and, if the instability threshold is overcome, they can be amplified. Due to the fact that the parallel pumping process always allows for the creation of counter-propagating waves with a vanishing sum of wavevectors, this gives access to the excitation of spin waves with large wavevectors, even with spatially extended pumping sources\cite{Serga:2014hz,Sandweg}. This way, parametric generation has enabled the excitation of short-wavelength spin waves in macroscopic systems, an important prerequisite for the observation of, for instance, the Bose Einstein condensation of magnons\cite{Mockel-2006-BEC,Clausen-2015-1}. Moreover, unlike conventional antennae\cite{Mockel2009, Schneider-2008-1, Pirro-2011-1}, it is not primarily sensitive to the net magnetic moment of the excited modes and, thus, their symmetry. In contrast, the wavevector dependence of the threshold fields is mainly determined by the above mentioned characteristics and, thus, a function of the ellipticity of the mode pairs and their relaxation frequency\cite{Kostylev-2007-PP,SW_basics}. Hence, parametric generation is a powerful tool to study the mode spectra in magnetic systems. Due to the fact that on the long run the amplification of the dominant mode suppresses all other modes, a knowledge of the dominant mode spectrum of the investigated structures is mandatory in order to design the optimum amplification of externally excited spin waves. In this Section, we will discuss the dominant mode spectra in the most commonly studied geometries in microstructures, namely in microscopic elliptical elements and longitudinally as well as transversely magnetized  spin-wave waveguides. We will show that in all these systems a favored generation of the fundamental mode can be achieved.

\subsection{Parametric generation in microstructured elliptical elements}
\label{Pgen-Ell}

\begin{figure}[t]
\center
{\caption{a) SEM micrograph of the investigated sample. b) Observed excitation spectra as a function of applied pumping power. c) Observed intensity distribution (left) and cross-section belonging to the peaks with detection frequency $f_0$, $f_1$ and $f_2$ (right). Frequencies are given in the detection frequency, which is equal to half of the pumping frequency $f_\mathrm{p}$ (Reprinted figures from H. Ulrichs et al., Parametric excitation of eigenmodes in microscopic magnetic dots, Phys. Rev. B. 84, 094401 (2011). Copyright (2016) by the American Physical Society.} 
 \label{Fig:Ulrichs}}
{\includegraphics[width=1.0\textwidth]{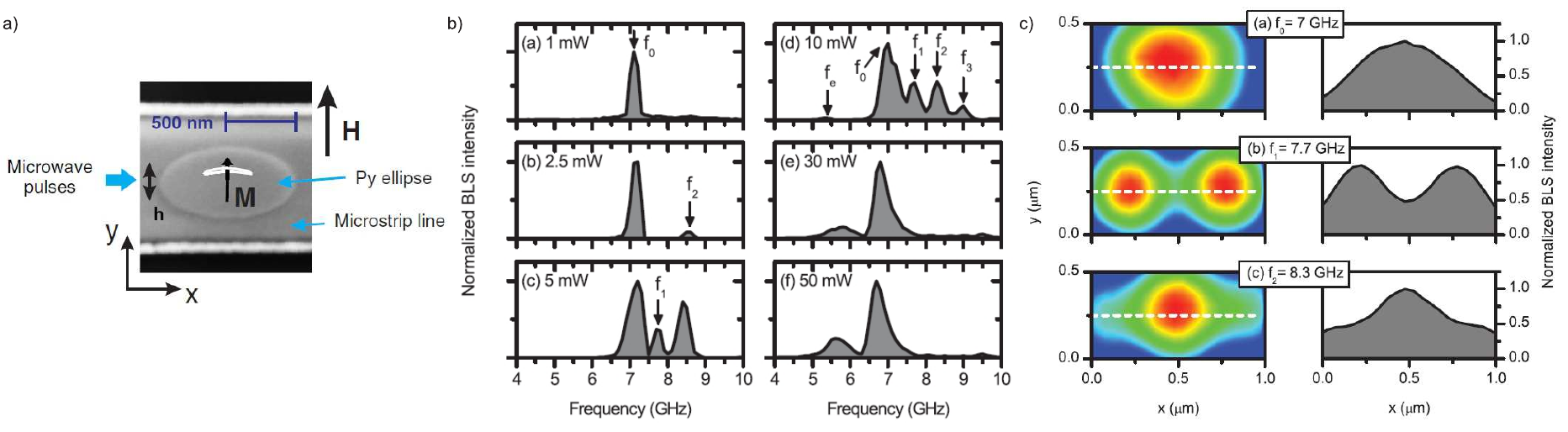}}
\end{figure}

Microstructured elliptical elements are widely used as basic elements in prototypes of magnetic random access memories\cite{MRAM1, MRAM2} and spin-transfer torque driven magnetic nano-oscillators\cite{ZengSTNO}. Hence, an understanding of their fundamental spin-wave spectrum, including odd as well as even modes, its relaxation characteristics and its coupling to an RF magnetic field are of vital importance for the optimization of these devices. In these microscopic elliptical elements, the boundaries give rise to a spectrum of discrete resonance frequencies associated with laterally standing spin-wave modes\cite{vogtvortex, Jorzdisc, Gubidisc, Stampdisc}. The parametric generation of these modes was studied in Ref. \cite{Ulrichs} by means of microfocussed Brillouin light scattering (BLS)\cite{BLS1,BLS2,Mock-1987-1} as well as in Ref. \cite{BobMcM} by means of ferromagnetic resonance force microscopy\cite{MRFM1, MRFM2}. In both of these works, magnetic elliptical elements have been fabricated on top of a micro-strip which provides the RF pumping field in the form of the Oersted field created by a microwave current in the strip. The configuration studied in Ref. \cite{Ulrichs} is depicted in Fig.~\ref{Fig:Ulrichs}~(a), where a scanning electron microscopy (SEM) micrograph of the investigated structure is shown. In this experiment, as well as in most other experiments regarding parametric generation and amplification in microscopic systems, the pumping field has been applied in pulses. This minimizes sample heating and the time the parametric amplifier is saturated. Moreover, it allows for a time-resolved investigation of the action of the pumping field on the spin-wave dynamics.

As has been demonstrated in Refs. \cite{Ulrichs} and \cite{BobMcM}, the discrete spectrum of standing modes give rise to the observation of distinct resonance peaks in the spectra of parametrically generated spin waves, which feature similar, yet slightly different, threshold powers. This can be comprehended from  Fig.~\ref{Fig:Ulrichs}~(b), where the BLS intensity is shown as a function of one-half of the applied pumping frequency, which corresponds to the detection frequency of the parametrically generated spin waves. For these measurements, the BLS laser has been positioned in the center of the elliptical element, which is magnetized along its short axis by an external bias field, and the frequency of the pumping field has been varied for various fixed peak powers of the microwave pumping field. At first, only one peak is observed ($1\,\mathrm{mW}$, frequency $f_0$). An increase of the pumping power leads first to an increased number of visible modes at higher frequencies ($2.5-10\,\mathrm{mW})$. At larger powers, nonlinear interactions lead to a broad excitation spectrum. An analysis of the spatial intensity distribution associated with the observed peaks at low powers yields the profiles of the corresponding modes. Figure~\ref{Fig:Ulrichs}~(c) shows space-resolved BLS maps of the mode profiles associated with the peaks $f_0$, $f_1$ and $f_2$. Hereby, it should be noted that the BLS intensity is directly proportional to the spin-wave intensity at a given position\footnote{BLS microscopy typically features a nominal spatial resolution of about $250\,\mathrm{nm}$\cite{BLS2}.}. The mode associated with the excitation with the lowest threshold is the fundamental mode $n = 1$, featuring one intensity maximum in the center of the dot. Since the higher modes feature an increasing number of nodes, their dynamic dipolar stray field and, consequently, their ellipticity of precession decreases. As a consequence, the higher modes feature a reduced coupling to the microwave pumping field. Since they additionally feature a higher frequency, they exhibit a larger relaxation rate (see Eq.~\ref{Eq:TauComp}). Hence, from Eq.~\ref{Eq:Threshold} it is expected that the first mode has the lowest parametric generation threshold, as it is visible in Fig.~\ref{Fig:Ulrichs}~(b). The peak that appears at the frequency $f_\mathrm{e} < f_0$ is associated with the parametric generation of an edge mode\cite{Bayer-2006,Jorzick-2001,Jorzick-2002}. Due to the fact that the changes in the relaxation frequency and in the coupling strengths are not very pronounced, the differences in the thresholds are rather subtle. Hence, parametric generation allows to excite modes of odd and even symmetry as well as edge modes with a comparable excitation strength and is well suited for the study of the entire mode spectrum in such microstructured elements. This makes this method an interesting alternative to conventional excitation schemes, which use a dynamic Oersted field perpendicular to the static magnetization\cite{Bailleul,Pirro-2011-1,Mockel2009}, since these cannot access even modes and whose excitation efficiency for higher modes decreases with $1/n$ due to the decreasing magnetic moment of the modes with increasing mode number.

\subsection{Parametric generation in longitudinally magnetized waveguides}
\label{Pgen-BV}

\begin{figure}[h]
\center
{\caption{a) Sketch of sample for the localized parametric generation in a longitudinally magnetized waveguide. b) Spin-wave wavevector as a function of the externally applied bias field for a fixed frequency of $f = 6\,\mathrm{GHz}$. The dashed lines represent the exchange-dominated branches of the spin-wave dispersion relations. c) Experimentally observed thresholds in comparison to theoretically expected thresholds (individually normalized). The shaded areas represent the dominantly generated mode. d) Representative profiles of the dominantly generated modes. (after \cite{Braecher-LocBV})} 
 \label{Fig:LocBV1}}
{\includegraphics[width=0.8\textwidth]{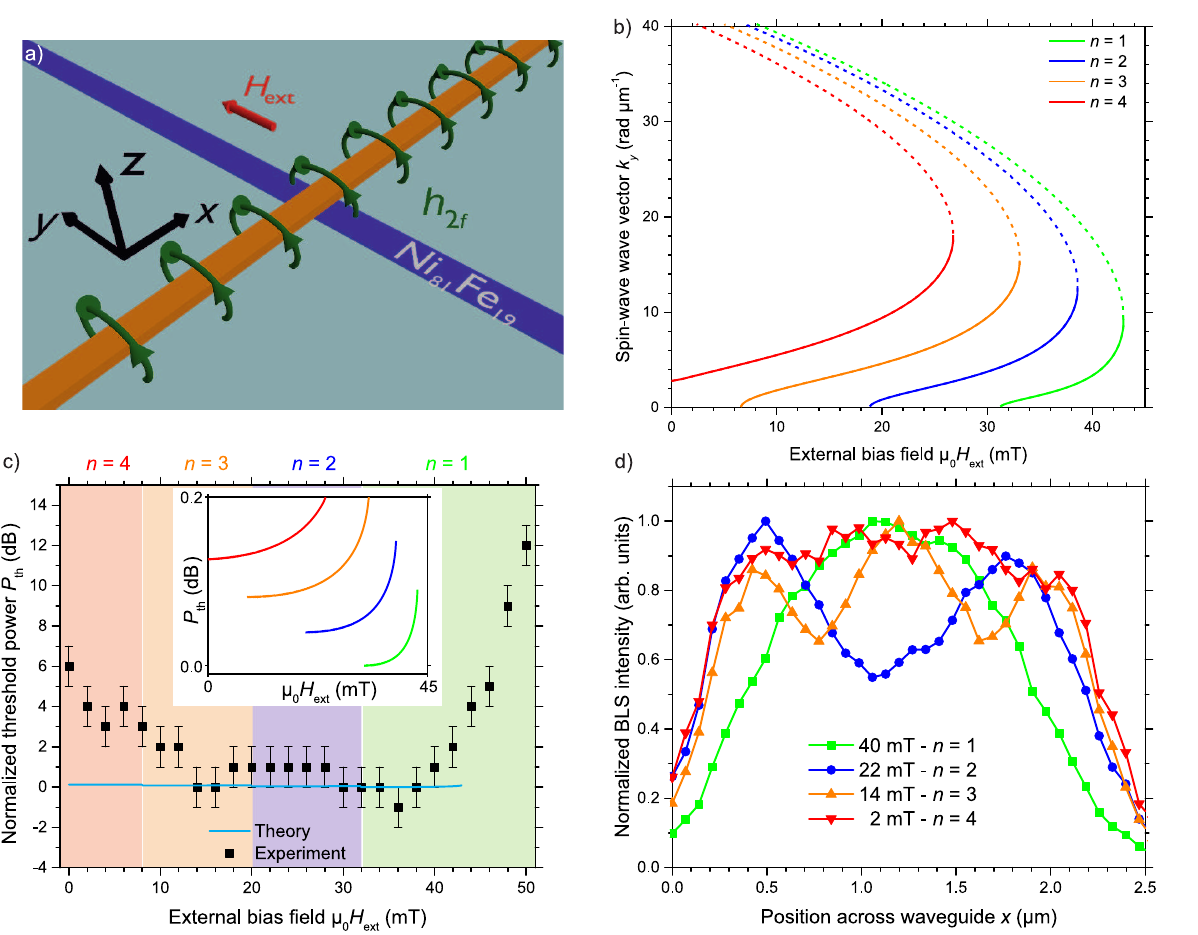}}
\end{figure}

In addition to its application as a tool to study or amplify the fundamental excitations in elliptical elements, the technique of parallel pumping is also well suited for the generation and amplification of spin waves in waveguides. As has been demonstrated in Ref. \cite{Braecher-LocBV}, a convenient way to realize a local spin-wave generation in a waveguide magnetized along its long axis can be realized by using a simple micro-strip-antenna. In a geometry like the one depicted in Fig.~\ref{Fig:LocBV1}~(a), where such an antenna is patterned on top of such a waveguide, the in-plane component of the AC Oersted fields provides the microwave pumping field interacting with the longitudinal component of the dynamic magnetization. Similar to the application to elliptical elements, this allows for the generation of odd and even spin-wave modes in the waveguide. In contrast to the out-of-plane component of the antenna field, which is the dominant conventional means of spin-wave excitation by a direct torque in this geometry, the localization of the in-plane component results in a local spin-wave excitation by parametric generation. Moreover, due to the creation of spin-wave pairs with $\pm k_y$, such an excitation source can practically excite arbitrarily short wavelengths, independent of its width. 

For a better comparison to the experiment in Ref. \cite{Braecher-LocBV}, in which the pumping frequency is fixed while the externally applied magnetic field is varied, Fig.~\ref{Fig:LocBV1}~(b) shows the spin-wave wavevector of the first four waveguide modes $n = 1$ to $n = 4$ as a function of the externally applied bias field for a fixed spin-wave frequency of $6\,\mathrm{GHz}$. For the calculations, a $2.2\,\mu\mathrm{m}$ wide, $20\,\mathrm{nm}$ thick waveguide made from Ni$_{81}$Fe$_{19}$ has been assumed, as has been investigated in Ref. \cite{Braecher-LocBV}. They have been performed assuming a saturation magnetization of $M_\mathrm{s} = 800\,\mathrm{kA}\,\mathrm{m}^{-1}$, an exchange constant of $A_\mathrm{ex} = 13\,\mathrm{pJ}\,\mathrm{m}^{-1}$ and an effective waveguide width of $2.3\,\mu\mathrm{m}$\cite{Guslienko-eff-width}. In the figure, the solid lines correspond to the dipolar branch of the dispersion relation and the dashed lines to the exchange branch. It is important to note, that for each field, the lowest available waveguide mode features the lowest available wavevector in the dipolar branch. Hence, it is supposed to feature the largest ellipticity and, consequently, the largest coupling to the pumping field. 

Figure~\ref{Fig:LocBV1}~(c) shows the experimentally determined thresholds of parametric generation together with the theoretically expected threshold powers. To achieve parametric generation, $10\mathrm{ns}$ long pumping pulses with a carrier frequency of $f_\mathrm{p} = 12\,\mathrm{GHz}$ have been applied to the antenna. The theoretical threshold powers have been calculated following Eqs.~\ref{Eq:TauComp} and \ref{Eq:V} by using Eq.~\ref{Eq:Threshold} and $P_\mathrm{th} \propto (\mu_0 h_\mathrm{th})^2$, with the same parameters used to derive the spin-wave dispersion relation in Fig.~\ref{Fig:LocBV1}~(b) and a Gilbert damping parameter of $\alpha = 0.0115$. The calculated thresholds have been normalized to their minimum value at $k_y = k_{||} = 0$ for the mode $n = 1$, corresponding to $\mu_0 h_\mathrm{th} = 33.8\,\mathrm{mT}$. The experimental threshold values have been normalized to the nominal peak power $P = 11\,\mathrm{dBm}$ at $\mu_0 H_\mathrm{ext} = 32\,\mathrm{mT}$, corresponding to $k_y = 0$ for the mode $n = 1$. The shading represents the observed dominant spin-wave mode, for which exemplary mode profiles are shown in Fig.~\ref{Fig:LocBV1}~(d). As can be seen from Fig.~\ref{Fig:LocBV1}~c), the modes $n = 1$ to $n = 4$ can be parametrically generated with a similar threshold. In agreement with the theoretical prediction, the experimentally determined threshold is approximately constant within the experimental resolution in a large field range, while it increases at large fields when no resonant excitation is possible anymore. However, there is also a clear discrepancy from the simple analytical theory described in Section \ref{Theoback} at low fields. This less accurate description is likely mediated by the fact that in this field-range the assumption $\mu_0 h_\mathrm{dyn} \ll \mu_0 H_\mathrm{stat}$, i.e., that the dynamic fields are much smaller than the static effective fields, is not satisfied anymore. Nevertheless, the analytical formalism is able to reproduce the observed dominant mode, as is evident in the inset of Fig.~\ref{Fig:LocBV1}~(c), where the threshold powers are shown with an adequate scale. As expected, the lowest available waveguide mode features the lowest expected threshold, a finding also predicted in Ref. \cite{Kostylev-2007-PP}. As can be comprehended from comparing Figs.~\ref{Fig:LocBV1}~(b) and (c), the changes of the dominant mode are associated with the respective origin of the dipolar dispersion relations where $k_{y,n} \approx 0$. The resonance field and frequency of these modes as well as their frequency-splitting can be tuned by changing the geometric extents of the spin-wave waveguide. Hence, parametric generation is a viable means to excite propagating spin waves in longitudinally magnetized waveguides, where the dominant mode can be tuned to be the desired odd or even waveguide mode by adjusting the geometry and the frequency and field combination. It should be noted that the jumps from one waveguide mode to another are analogous to the transitions between the different standing modes across the film thickness in thin, but not ultra-thin, films (cf., e.g., Refs. \cite{Kostylev-1995-1,Vendik-1977-PP,Kalinikos-1989-PP, Kalinikos-1985-PP,Kalinikos-1985-PP2,Wiese-1994}).

\begin{figure}[t!]
\center
{\caption{a) Spin-wave dispersion relation of the first three waveguide modes in an effectively $600\,\mathrm{nm}$ wide, $20\,\mathrm{nm}$ thick Ni$_{81}$Fe$_{19}$ waveguide in the absence of external bias fields. b) Excitation spectrum for various microwave powers at zero applied bias-field. c) Experimentally determined spin-wave decay length (black squares) in comparison to theoretical expectation (blue line) (after \cite{Phd-Braecher}).} 
 \label{Fig:LocBV2}}
{\includegraphics[width=1.0\textwidth]{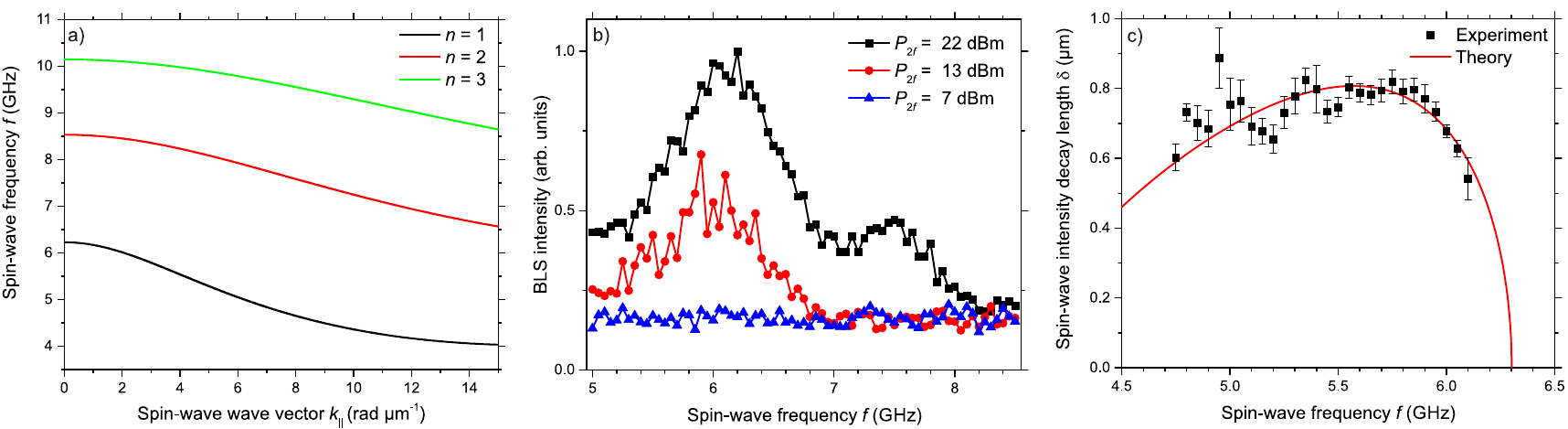}}
\end{figure}
A particularly interesting application of this excitation geometry is the study of waveguides in their remanent state, i.e., in the absence of external bias fields. This paves the way for magnonic applications without the need for an applied magnetic field and allows, for instance, to study the interaction of spin waves with domain walls in the remanent state\cite{Pirro-2015-1, DW1, DW2, DW3}. As mentioned above, by varying the dimensions of the waveguide or the pumping frequency, the fundamental waveguide mode $n = 1$ can be shifted to be the dominant mode at zero field. This is demonstrated in Fig.~\ref{Fig:LocBV2}~(a), where the calculated spin-wave dispersion relations of the first three waveguide modes in a $600\,\mathrm{nm}$ wide and $20\,\mathrm{nm}$ thick Ni$_{81}$Fe$_{19}$ spin-wave waveguide are shown in the absence of an external bias field (material parameters as stated above). Due to the small width of the waveguide, the splitting between the adjacent waveguide modes is fairly large and, consequently, each mode exists basically in a different frequency range. Thus, by changing the working frequency, the desired mode can be chosen. 

Figure \ref{Fig:LocBV2}~(b) shows the parametric excitation spectrum as a function of frequency in the absence of an external bias field in a waveguide with these dimensions for different pumping powers. The parametric generation has been performed by means of a $1.2\,\mu\mathrm{m}$ wide micro-strip antenna. If the threshold is exceeded, spin waves are excited in a wide frequency and, thus, wavevector range. Hereby, the excitation with the lowest threshold is associated with the generation of the fundamental waveguide mode $n = 1$. According to the spin-wave dispersion, the smallest excited wavelength in this scenario corresponds to a spin-wave wavelength of ca. $700\,\mathrm{nm}$. This wavelength is below the linear excitation range of the micro-strip-antenna, which is limited to wavelengths larger than the antenna width. This demonstrates the feasibility of parametric generation to excite a large wavevector range. Figure \ref{Fig:LocBV2}~(c) shows the exponential spin-wave decay lengths extracted from the space-resolved BLS intensity scanning away from the antenna\cite{Stancil,Pirro-2011-1}. The experimentally determined decay lengths are compared to the theoretically estimated decay length $\delta = v_\mathrm{g}/\omega_\mathrm{r}$ derived from Eqs.~\ref{Eq:Disp-tf} and \ref{Eq:TauComp}, which is represented by the red solid line. As can be seen from the figure, there is a good agreement between the observed and expected spatial decay of the waves emitted from the antenna, underlining nicely the propagating character of the excitation. For larger powers, the threshold for the parametric generation of the mode $n = 2$ is overcome and, consequently, the generation of this mode is observed for frequencies $f \gtrsim 7\,\mathrm{GHz}$ in a large wavevector range as well. 

\subsection{Parametric generation in transversely magnetized waveguides}
\label{Pgen-DE}

\begin{figure}[b!]
\center
{\caption{a) Spin-wave wavevector as a function of the effective field for the first 10 waveguide modes for a fixed spin-wave frequency of $f = 6\,\mathrm{GHz}$ in a transversely magnetized Ni$_{81}$Fe$_{19}$ waveguide. b) Corresponding threshold fields.} 
 \label{Fig:GenDE1}}
{\includegraphics[width=0.8\textwidth]{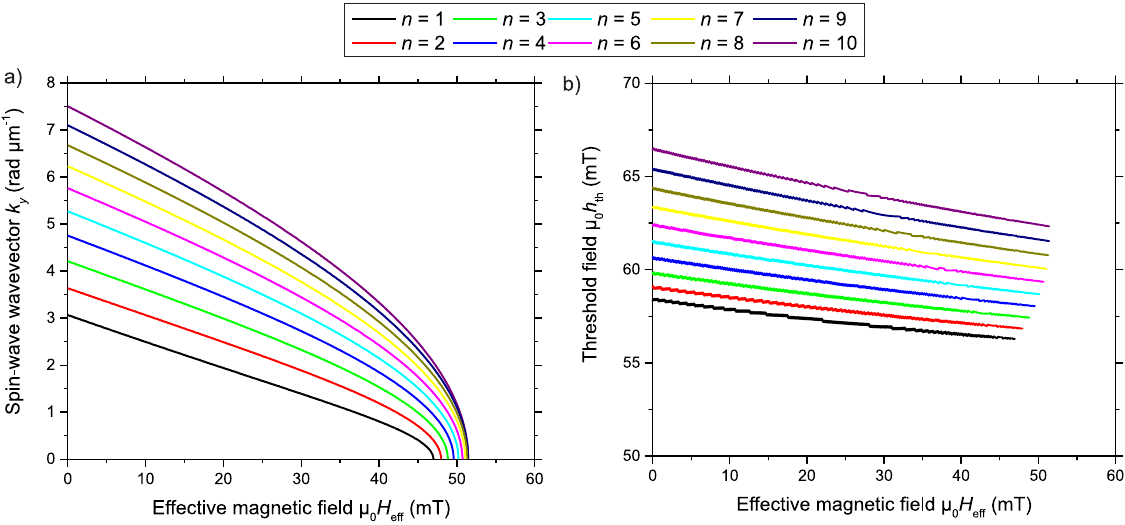}}
\end{figure}
Due to the larger group velocity\cite{SW_basics, Stancil} and the large sensitivity of the spin-wave spectrum for the interface conditions\cite{Burkard-PRL-1990}, spin waves in transversely magnetized waveguides are of great interest for the realization of magnonic devices. This geometry, commonly also referred to as the Damon-Eshbach geometry, is the geometry in which most experiments in microstructured spin-wave waveguides have been performed\cite{Mockel2009,Mockel2011,Pirro-2011-1,Mockel-2015,BLS2,Koji-2010,Koji-2012}. Figure~\ref{Fig:GenDE1}~(a) shows the calculated spin-wave wavevector as a function of the effective magnetic field in a $w_\mathrm{eff} = 2.25\,\mu\mathrm{m}$ wide, $40\,\mathrm{nm}$ thick Ni$_{81}$Fe$_{19}$ spin-wave waveguide for the first 10 waveguide modes for a fixed frequency of $f = 6\,\mathrm{GHz}$. As can be seen from the figure, the separation between these modes is much less pronounced than in a longitudinally magnetized waveguide. Nevertheless, the lowest available waveguide mode features the lowest threshold of all modes, as can be comprehended from Fig.~\ref{Fig:GenDE1}~(b), where the corresponding calculated threshold fields are shown assuming the same parameters and a Gilbert damping parameter of $\alpha = 0.008$. Hence, also in transversely magnetized waveguides, the fundamental waveguide mode can be tuned to be the dominantly amplified spin-wave mode. This has been experimentally verified in Ref. \cite{Braecher-2011-1} by a study of the parametric generation in such a Ni$_{81}$Fe$_{19}$ waveguide. 

\begin{figure}[b!]
\center
{\caption{a) Sample layout for the homogeneous parametric generation in a transversely magnetized spin-wave waveguide. b) Color-coded BLS intensity at $f = 6\,\mathrm{GHz}$ for a fixed pumping frequency of $f_\mathrm{p} = 12\,\mathrm{GHz}$ as a function of the applied bias field and pumping peak power. c) Exemplary measured mode profiles at $f = 6\,\mathrm{GHz}$ for different bias fields (after \cite{Braecher-2011-1}).} 
 \label{Fig:GenDE2}}
{\includegraphics[width=0.8\textwidth]{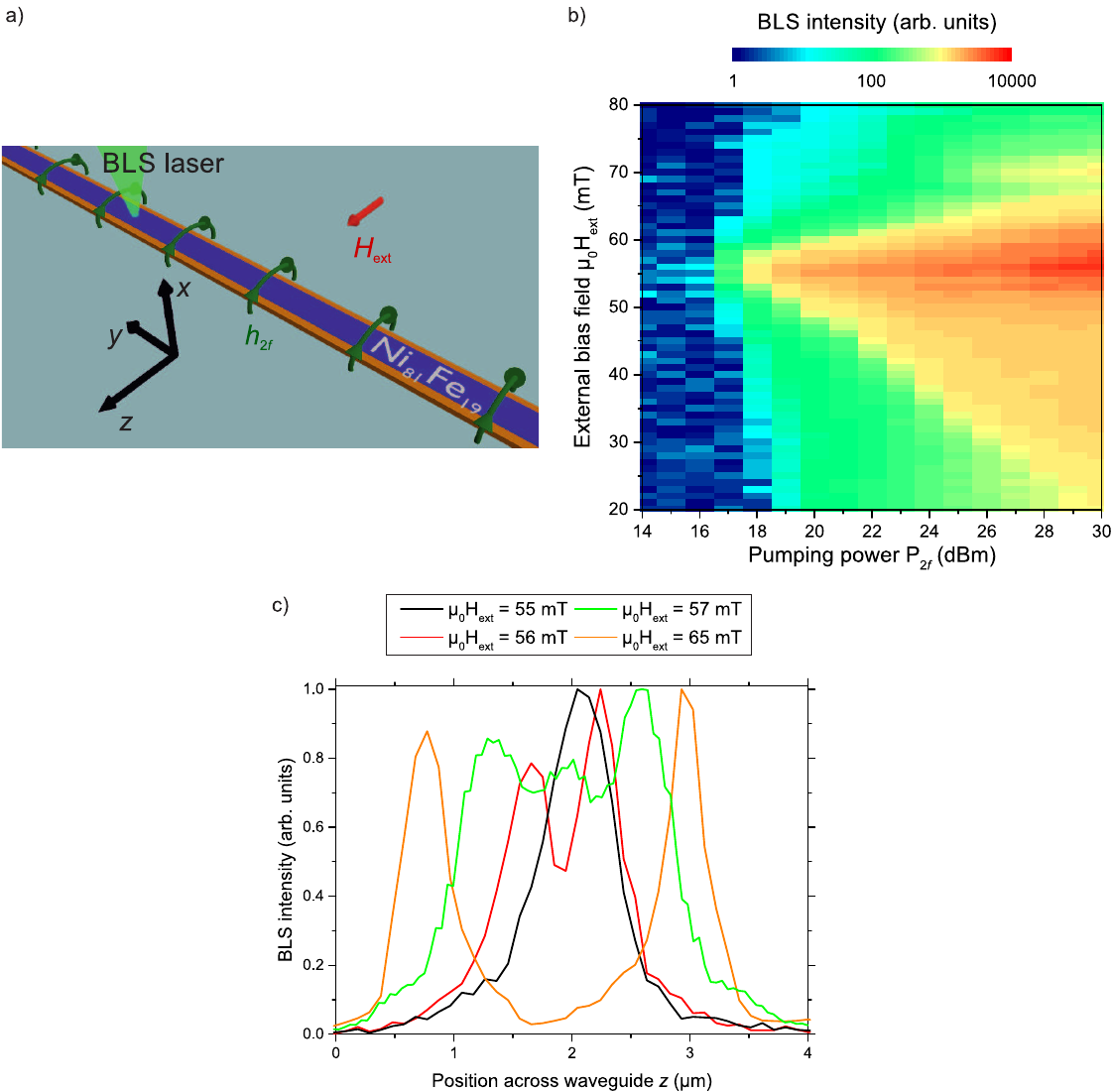}}
\end{figure}
Using the sample layout depicted in Fig.~\ref{Fig:GenDE2}~(a), the Oersted field created by a microwave current flowing through the micro-strip underneath the transversely magnetized Ni$_{81}$Fe$_{19}$ waveguide provides the pumping field interacting with the dynamic magnetization. This way, a homogeneous parametric generation along the waveguide could be achieved, similar to the homogeneous parametric generation in the microscopic elliptical elements discussed in Section \ref{Pgen-Ell}. Figure~\ref{Fig:GenDE2}~(b) shows the color-coded BLS intensity at $f = 6\,\mathrm{GHz}$ as a function of the externally applied bias field ($y$-axis) and the pumping microwave peak power ($x$-axis). In the experiment, the microwave pumping has been applied in $100\,\mathrm{ns}$ long pulses with a carrier frequency of $f_\mathrm{p} = 12\,\mathrm{GHz}$. As can be seen, parametric generation can be observed in a wide range of magnetic fields, again with no large variation in the threshold power. Figure~\ref{Fig:GenDE2}~(c) presents the observed profiles of the dominantly generated modes. Below an external field of $\mu_0H_\mathrm{ext} = 56\,\mathrm{mT}$, which can be associated with an effective field of $\mu_0H_\mathrm{eff} = 48\,\mathrm{mT}$ within the waveguide\cite{Bayer-2006}, the fundamental waveguide mode $n = 1$ is generated, as is expected from Fig.~\ref{Fig:GenDE1}~(b). For $\mu_0H_\mathrm{ext} = 56\,\mathrm{mT}$, the second waveguide mode is the lowest available waveguide mode and, thus, dominantly amplified. A further increase of the field leads to a further increase of the mode number until at magnetic fields exceeding $\mu_0H_\mathrm{ext} \approx 59\,\mathrm{mT}$, no more waveguide modes can be resonantly excited. Consequently, at higher fields one observes the parametric generation of edge modes. Thus, also in extended magnonic waveguides, parametric generation gives access to the excitation of odd and even spin-wave modes with finite wavevector. In particular, the fact that the dominantly generated waveguide mode is the fundamental waveguide mode in a large wavevector range is the basis of the large potential for the application of parallel pumping in microstructures, as will be discussed in the following Sections. However, it should be noted that due to the very small differences in the threshold fields for waveguides with widths in the $\mu\mathrm{m}$ range, small changes in the spin-wave lifetime due to wavevector dependent damping\cite{SW_basics,Stancil,Krivosik:2010tz,Heusler-PRL,Kosytlev-Eddy} or changes of the dispersion, such as avoided crossings due to dipolar interaction between the waveguide modes\cite{Grigo2013,Chumak-2009-MC,Kostylev-2007-Disp}, can render the generated mode spectra more complicated.

\section{Amplification of traveling waves}
\label{Pamp}

One of the main applications of parallel pumping of spin waves is the amplification of externally excited, traveling spin-wave packets carrying information in form of their amplitude, phase or both. In the following, these waves are referred to as signal spin waves. The technique of parallel pumping has been successfully applied to traveling spin-wave packets in macroscopic spin-wave waveguides made from YIG. Among others, the amplification of solitons\cite{Kabos}, the cloning and trapping of spin-wave packets\cite{Vitaliy}, as well as the storage and recovery in the spin-wave system\cite{Schaefer-2011,Chumak-2009-1,Serga-2007} have been realized on the macroscopic scale. However, in these systems, the parametric generation of the dominant spin-wave group, which is generally not the externally excited, signal-carrying spin wave, always leads to a suppression of the signal spin waves after a finite time. Hence, the phase-information carried by the signal spin waves is lost. The aforementioned fact, that the preferably generated spin-wave mode in microscopic spin-wave waveguides is the fundamental waveguide mode is a new paradigm for the applicability of parallel pumping for the spin-wave amplification. Since, consequently, in these systems the externally excited, signal carrying spin wave coincides with the dominantly generated spin-wave mode, a particularly efficient parametric amplification should be possible. Due to the beneficial properties of parametric amplification, such as frequency- and mode-selectivity and the conservation of the spin-wave phase, it has several advantages over alternative methods, such as the amplification via the STT effect induced by an incident spin current, which lack such selectivity. In the following Section, the potential and the limits of parallel pumping as a mere means of spin-wave amplification are presented, highlighting the role of the parametric generation, which, at room temperature, is still interfering with the amplification of the externally excited spin waves. It will be shown that its role is quite different from macroscopic systems: The fact that the same wave as the signal mode is predominantly generated means that its generation is not generally suppressing the signal wave. However, the fact that the thermal spin waves with exactly the same spin-wave frequency and wavevector are amplified gives rise to two problems: 1. It can lead to an undesired creation of \textit{noise}, since - at least for a detection only sensitive to the spin-wave intensity - the generated waves cannot be distinguished from the signal waves. 2. One feature of the parametric amplifier is its saturation after a certain spin-wave intensity is reached within the amplifier (cf. Section \ref{Theo-PPA-Sat}). Thus, as will be discussed in the following, the parametric generation can lead to an undesired \textit{saturation} of the amplifier before the traveling signal spin waves arrive. 

\subsection{Local spin-wave amplification and time- and power-dependence of the parametric amplification}
\label{Pamp-Loc}

\begin{figure}[b!]
\center
{\caption{a) Sample layout for the localized parallel parametric amplification studied by BLS microscopy. The actual amplifier is defined by a geometric constriction, resulting in a locally enhanced pumping field. b) Space-resolved BLS intensity arising from the parametric amplification of the externally excited spin waves (black squares), from the antenna excitation only (red dots) and from parametric generation only (green triangles). The dashed blue line shows the intensity arising from the sum of the individual intensities arising from the external spin-wave excitation and the parametric generation. The shading marks the position and the length of the geometrical constriction. c) Amplification measured $25\,\mu\mathrm{m}$ from the amplifier. d) Time-resolved BLS intensity measured behind the amplifier for different delays between the pumping pulse and the spin-wave packet. The shading represents the temporal extent of the spin-wave packet. (Reprinted from T. Br\"{a}cher et al., Time- and power-dependent operation of a parametric spin-wave amplifier, Appl. Phys. Lett. 105, 232409 (2014) with the permission of AIP publishing)} 
 \label{Fig:Pamp-Loc1}}
{\includegraphics[width=0.8\textwidth]{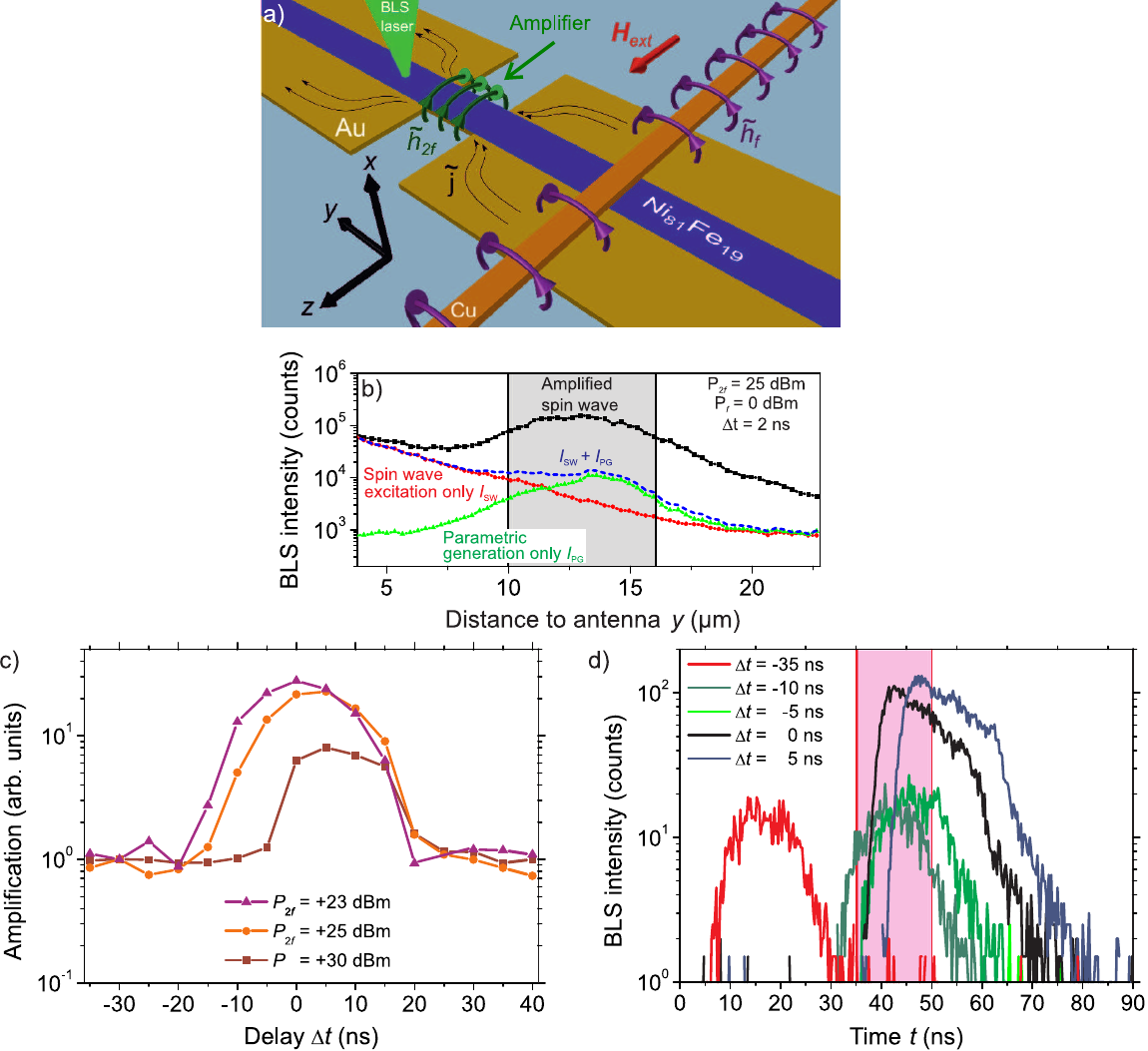}}
\end{figure}
Figure \ref{Fig:Pamp-Loc1}~(a) shows the layout of a local spin-wave amplifier which has been experimentally investigated in Ref. \cite{APL-2014-3}. A spin-wave waveguide has been patterned on top of a pumping micro-strip. In addition, a micro-strip-antenna has been patterned across the spin-wave waveguide to allow for a local and coherent spin-wave excitation at frequency $f_\mathrm{s}$. Similar to the experimental realization for the parametric generation, a microwave with frequency $f_\mathrm{p} = 2 f_\mathrm{s}$ provides the microwave pumping field for the parametric amplification. The spin-wave excitation and the parallel pumping field are applied in pulses. In addition to the aforementioned minimization of heating, this also allows to study the influence of the relative timing of the amplification with respect to the arrival of the spin-wave packet within the amplifier. This is experimentally controlled by the delay between the arrival of the externally excited spin wave packet and the application of the pumping pulse $\Delta t$, where $\Delta t < 0$ implies that the pumping starts before the spin-wave packet arrives.

The amplifier shown in Fig.~\ref{Fig:Pamp-Loc1}~(a) features a general width of $20\,\mu\mathrm{m}$, which is reduced to a width of $4\,\mu\mathrm{m}$ over a length of $6\,\mu\mathrm{m}$ at a distance of $10\,\mu\mathrm{m}$ from the antenna. The larger current density in this region with smaller cross-section provides a locally enhanced microwave Oersted field. The combination of this local enhancement with the threshold character of parametric amplification results in the actual localization of the parametric spin-wave amplifier\cite{APL-2014-1}. In the experiment in Ref. \cite{APL-2014-3}, the spin-wave excitation and the parallel pumping have been applied in $15\,\mathrm{ns}$ and $25\,\mathrm{ns}$ long microwave pulses, respectively.  The microwaves at $f_\mathrm{s}$ and $2f_\mathrm{s}$ have been created by two independent microwave sources. The output of the amplifier has been studied as a function of the power of the pumping as well as the delay $\Delta t$ between the two pulses. Consequently, the time-integrated spin-wave intensity behind the amplifier, which has been measured by BLS microscopy, can be constituted of the traveling spin-wave packet excited by the antenna, a packet of parametrically generated waves and, if the parallel pumping interacts with the traveling waves, a packet of amplified signal spin waves. 

The spatial evolution of the spin-wave intensity arising form the interplay of an externally excited, propagating spin-wave packet with such a localized parallel pumping field is shown in Fig.~\ref{Fig:Pamp-Loc1}~(b) by the black squares. For the applied moderate pumping power of $P_{2f} = 25\,\mathrm{dBm}$ and for an arrival of the spin-wave packet within the amplifier shortly before the pumping pulse is applied ($\Delta t = 2\,\mathrm{ns}$), a strong spin-wave amplification is observed. This can be comprehended by a comparison with the dashed blue line, which is the sum of the intensity arising from the antenna excitation only (red dots) and the intensity arising from parametric generation only (green triangles). In contrast to the BLS intensity arising from the antenna excitation only, the amplifier is able to restore the spin-wave amplitude to its initial intensity level in the vicinity of the antenna. This way, it greatly enhances the propagation distance of the traveling waves, which, in the absence of parallel pumping, are found to decay exponentially with a decay length on the order of a few microns until their intensity falls below the experimental noise level of the BLS microscope. The green triangles demonstrate that such a local parametric amplifier can also be used as a source for local parametric generation, an aspect which has been studied in more detail in Ref. \cite{APL-2014-1}.

Figure \ref{Fig:Pamp-Loc1}~(c) shows the output of the amplifier measured at a distance of $25\,\mu\mathrm{m}$ from the antenna. The output has been normalized to the intensity measured at a delay of $-30\,\mathrm{ns}$, i.e., for a start of the pumping pulse $30\,\mathrm{ns}$ before the traveling spin-wave packet arrives. In this case there is no temporal overlap between the two pulses, no parametric amplification is observed and the output is simply given by the sum of the parametrically generated spin-wave packet and the incoming spin-wave packet, i.e., it corresponds to the blue dashed line in Fig.~\ref{Fig:Pamp-Loc1}~(b). This way, the output is a measure of the amplification, defined by the ratio of the output of the amplifier for the interaction of the signal spin waves with the pumping field over the output obtained if there is no interaction, i.e., no temporal overlap. It should be noted that in the latter case, the spin-wave packet excited at the antenna has already traveled over a distance of $25\,\mu\mathrm{m}$. Consequently, its intensity has already decayed significantly and, in fact, the main contribution to the measured intensity for a delay of $-30\,\mathrm{ns}$ shown in Fig.~\ref{Fig:Pamp-Loc1}~(c) arises from the generated packet. This can also be comprehended from Fig.~\ref{Fig:Pamp-Loc1}~(b), since the signal spin-wave intensity in the absence of parametric amplification is already below the noise level of the BLS at $y = 25\,\mu\mathrm{m}$. As the delay is shifted and the pulses overlap, a pronounced increase of the BLS intensity is observed, which is a fingerprint of the parametrically amplified spin waves. As can be seen from Fig.~\ref{Fig:Pamp-Loc1}~(c), the strength and the delay-dependence of the amplification depend strongly on the applied peak power of the pumping pulse. For a comparably low pumping power of $23\,\mathrm{dBm}$, which exceeds the parametric instability threshold by $3\,\mathrm{dB}$, the largest amplification by a factor of about $30$ is achieved and the amplification is rather symmetric as a function of the delay. As the pumping power is increased, the amplification reduces and the delay-dependence becomes increasingly asymmetric. At high powers, such as $30\,\mathrm{dBm}$, the curve is strongly asymmetric and a notable amplification is only observed, if the spin-wave packet enters the amplifier before the pumping starts (delay $\Delta t \geq 0$). 

This delay- and power-dependence are caused by the interaction of the pumping field with the thermal spin waves. The fact that the overall amplification decreases is due to the fact that the amplifier saturates very fast at large pumping powers far beyond the instability threshold. Consequently, the relative number of parametrically generated waves in comparison to amplified signal waves increases strongly with increasing pumping power. Consequently, the noise, i.e., the amount of undesired parametrically generated waves, created by the amplifier increases and the overall amplification drops according to the definition given above. 

To comprehend the observed dependence on the delay at large pumping powers, it is instructive to analyze the BLS intensity as a function of time. Figure \ref{Fig:Pamp-Loc1}~(d) shows the time-resolved BLS intensity arising for different delays at a pumping power of $30\,\mathrm{dBm}$. The shading represents the temporal extents of the spin-wave packet, whose temporal position is not shifted during the experiment. Due to the aforementioned strong decay of the traveling spin waves, the externally excited spin-wave packet is not visible in the time-resolved measurement if it is not amplified by the pumping field. The red curve shows the BLS intensity arising for a negative delay of $-35\,\mathrm{ns}$, i.e., for a complete displacement of the spin-wave packet and the pumping pulse in time. The detected spin-wave intensity is, thus, due to parametric generation only. After a initial fast increase of the BLS intensity, the intensity saturates quickly due to the nonlinear interaction of the generated spin waves and the pumping field. The parametrically generated packet is easily visible, due to the strong applied pumping field. In analogy to the time resolved measurements, the observed time-integrated intensity behind the amplifier at large pumping powers is (cf. Fig.~\ref{Fig:Pamp-Loc1}~(c)), thus, as mentioned before, mainly originating from parametric generation for this delay.

As can be seen from the green curves, even for a substantial overlap of the spin-wave packet with the pumping pulse, the time-dependent output of the amplifier is unchanged, as well as the time-integrated intensity. This is due to the fact that the parametric generation has already saturated the amplifier before the traveling signal spin-wave packet arrives. Hence, no amplification of the externally excited packet is observed and the measured time-integrated intensity is still only due to parametric generation.

Only if the spin-wave packet arrives first (delay $\Delta t \geq 0$, black and blue curve), the shape of the output pulse is changed and an amplification of the spin-wave packet in time takes place. Consequently, the time-integrated output changes and a net amplification is observed. This highlights the aforementioned two effects of parametric generation in microstructures: It can lead to the creation of undesired background noise and to an unwanted saturation of the amplifier. Therefore, in particular for applications where a strong pumping is needed to achieve a fast operation and large output powers, it is mandatory that the pumping starts after the signal spin waves have arrived at the amplifier. Moreover, short pumping pulses should be used to minimize the time the amplifier is saturated. In contrast, at low pumping powers, where the increase of the spin-wave intensity with time is slower (cf. Eq.~\ref{Eq:Intvicth}) and it takes a longer time until saturation is reached, the delay between the signal spin-waves and the pumping becomes less important. In this case the delay-dependence is not pronounced and a substantial amplification can be achieved, even for negative delays (cf. violet upward triangles in Fig.~\ref{Fig:Pamp-Loc1}~(c)). As will be discussed in the following Subsection, these findings have important implications for the global amplification of traveling waves.

\subsection{Global spin-wave amplification}
\label{Pamp-glob}

\begin{figure}[h]
\center
{\caption{a) Sample layout for the global parallel parametric amplification studied by BLS microscopy. b) BLS intensity scanning away from the micro-strip antenna. Blue: Interplay of pumping field with externally excited signal spin waves. Green: Only parametric generation. Violet: Only spin-wave excitation (after \cite{Phd-Braecher}).} 
 \label{Fig:Pamp-glob1}}
{\includegraphics[width=0.8\textwidth]{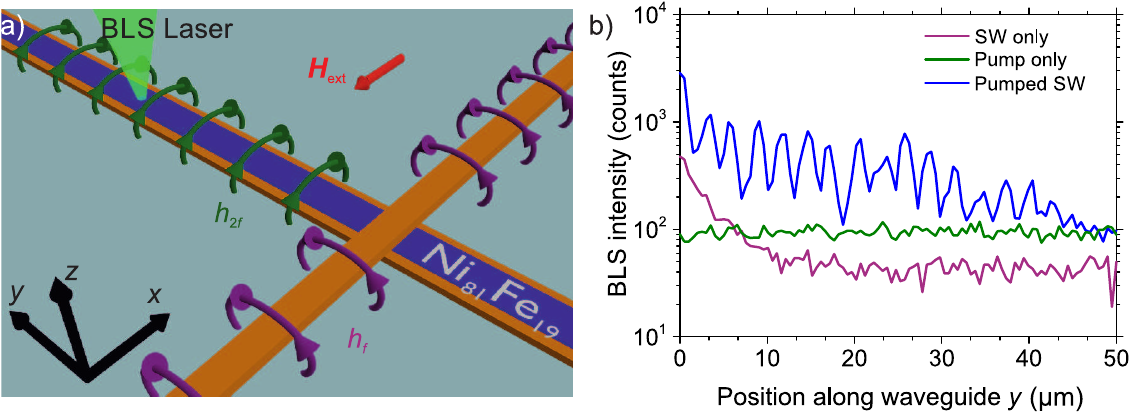}}
\end{figure}

For many applications, it is desired to compensate the spin-wave damping in the entire magnonic network. The technique of parallel parametric amplification is well suitable to achieve this aim. However, for such a \textit{global} parametric amplifier, parametric generation can impose limitations on the overall functionality\cite{APL-2014-2}, which will be discussed in this Subsection. The layout of an exemplary global parametric amplifier for a transversely magnetized magnonic waveguide is schematically shown in Fig.~\ref{Fig:Pamp-glob1}~(a). Similar to the parametric generation in a spin-wave waveguide, a micro-strip under the waveguide provides the microwave pumping field at frequency $f_\mathrm{p} = 2 f_\mathrm{s}$. An additional micro-strip-antenna patterned across the waveguide provides the local, coherent spin-wave excitation at frequency $f_\mathrm{s}$. 

A direct way to distinguish the amplified signal spin wave from the parametrically generated waves can be realized by a phase-locking of the spin-wave excitation and the microwave pumping. This can be achieved technically by generating them mutually using a single microwave source, whose output is split and frequency-doubled in one of the microwave paths. By the use of a phase-shifter in one of the microwave paths, the relative phase between the signal spin waves and the microwave pumping field can be shifted. Due to the fact that the parallel pumping process creates pairs of counter-propagating waves with a fixed phase relation determined by Eq.~\ref{Eq:Phaserel}, the amplification of a coherent wave with a well-defined phase results in a stationary interference pattern with a spatial periodicity of one half of the spin-wave wavelength. In contrast, the amplification of thermal waves results in the formation of a spatially homogeneous intensity distribution. This way, the amplification of thermal and signal waves can be distinguished directly by analyzing the stationary intensity distribution. An example of this is shown in Fig.~\ref{Fig:Pamp-glob1}~(b), where the stationary intensity distribution along the waveguide (measured along the waveguide center-axis) is shown for only the spin-wave excitation, only parametric generation and for the interplay of the parallel pumping field with the externally excited spin waves. The spin-wave excitation and the pumping field have been provided by $20\,\mathrm{ns}$ long microwave pulses with a carrier frequencies of $f = 7.13\,\mathrm{GHz}$ and $f_\mathrm{p} = 14.26\,\mathrm{GHz}$, respectively. The waveguide made from Ni$_{81}$Fe$_{19}$ features a geometrical width of $4.2\,\mu\mathrm{m}$ and a thickness of $40\,\mathrm{nm}$ and is magnetized along its short axis by an applied external bias field of $50\,\mathrm{mT}$. The microwave pulse have either been applied independently or, in case of the demonstration of parametric amplification, the spin-wave packet has been launched a few nanoseconds before the start of the pumping pulse. 
 
As can be seen from the figure (blue curve), the interplay of the pumping field with the traveling spin waves leads to a compensation of the otherwise significant decay of the traveling waves (cf. violet curve) and gives rise to a spatial modulation of the spin-wave intensity. This spatial modulation is the standing wave pattern which results from the interference of the created counter-propagating signal and idler waves with the same frequency $f_\mathrm{s}$ but featuring opposite wavevectors $\pm k_\mathrm{s}$. In contrast, in the absence of a coherent spin-wave excitation, parametric generation leads to a constant intensity level (cf. green curve), since the phase of the parametrically generated magnons is random. Despite the compensation of the spin-wave damping, the intensity modulation in the case of parametric amplification diminishes for distances larger than $y \approx 40\,\mu\mathrm{m}$ as well and the intensity drops to the level of the parametric generation. The reason for this is analog to the delay dependence of the amplification in a local amplifier: For larger distances, the parametric generation has already locally saturated the amplifier before the traveling spin-wave packet arrives. 

\begin{figure}[b!]
\center
{\caption{a) Time-resolved BLS intensity showing the rising edges of the spin-wave intensity at different distances from the antenna. The red lines are guides to the eye and their intercept with the horizontal dashed line marks the extrapolation of the arrival time. b) Minimum propagation range as a function of the applied supercritical pumping power. c) Signal velocity as a function of the applied pumping power. In Series 1, the pumping starts at the same time the spin-wave packet is launched while for Series 2, it only starts a few nanoseconds after the spin-wave excitation (after \cite{Phd-Braecher}).} 
 \label{Fig:Pamp-glob2}}
{\includegraphics[width=1.0\textwidth]{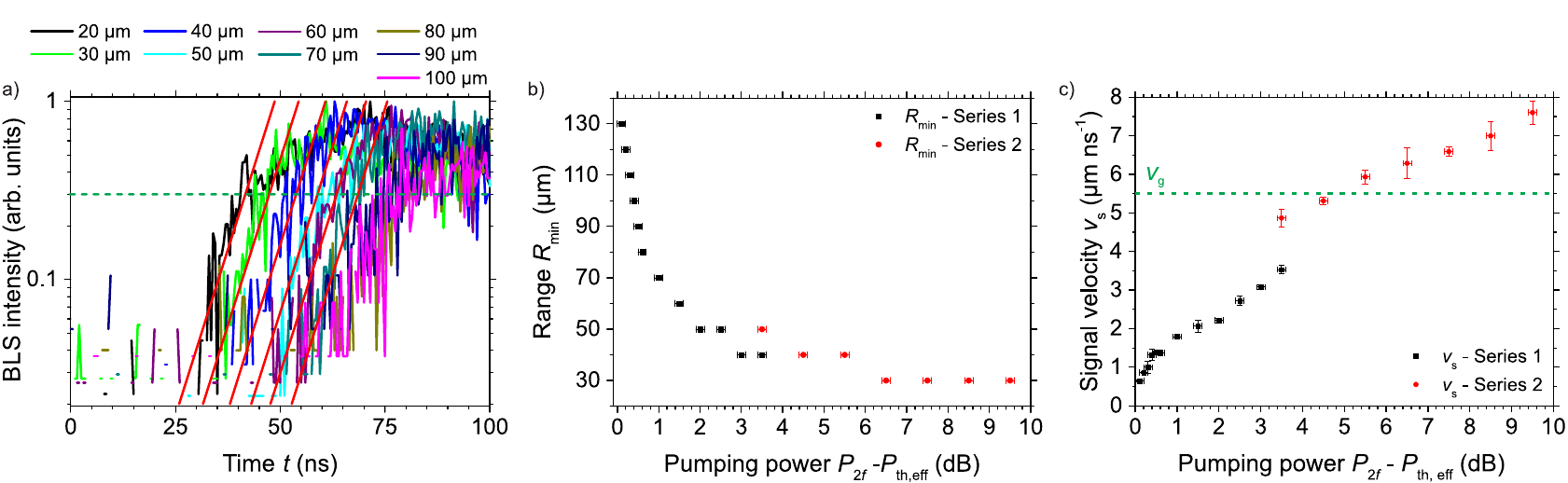}}
\end{figure}
To visualize this, Fig.~\ref{Fig:Pamp-glob2}~(a) shows the rising edge of the time-resolved BLS intensity measured at different distances from the antenna. For these measurements, the length of the pumping pulses has been increased to $200\,\mathrm{ns}$, the external field has been changed to $20\,\mathrm{mT}$ and the microwave frequency $f$ has been changed to $f_\mathrm{s} = 6.5\,\mathrm{GHz}$ (i.e. $f_\mathrm{p} = 13\,\mathrm{GHz}$). This results in a significantly smaller spin-wave wavelength of $\lambda = 2.3\pm0.4\,\mu\mathrm{m}$ in comparison to the waves excited in the measurements shown in Fig.~\ref{Fig:Pamp-glob1}. The long pumping pulses ensure a sufficient interaction time between the spin-wave packet and the pumping pulses. As can be seen from the figure, for distances below $80\,\mu\mathrm{m}$, the BLS intensity is emerging from the noise level at different times. This is a direct consequence of the amplification of the propagating waves, which need longer times to reach larger distances. However, for distances beyond $80\,\mu\mathrm{m}$, no further displacement is observed. In fact, the temporal evolution at a distance of $80\,\mu\mathrm{m}$ is identical with the temporal evolution of the parametric generation only. Thus, for larger distances the amplifier is already saturated prior to the arrival of the propagating waves. This is the reason for the fact that the stationary interference pattern in Fig.~\ref{Fig:Pamp-glob1}~(b) is only visible up to a certain point in space. Beyond that point in space, the amplifier is saturated by parametric generation and no standing interference pattern, which is a fingerprint of the amplification of the coherently excited signal spin waves, is observed. 

Time-resolved BLS measurements like the one shown in Fig.~\ref{Fig:Pamp-glob2}~(a) can be used to extract the propagation range of the traveling spin-wave packet\footnote{It is also possible to evaluate the range by analyzing the maximum distance at which the spatial modulation due to the interference of the signal and idler waves is observed. Both methods yield comparable results for the range.}: The maximum distance that the amplified wave front travels is at a distance in front of the point in space where the generation has already saturated the amplifier and beyond which no further displacement of the spin-wave intensity in time is observed. The minimum range of the traveling waves is, in return, associated with the last point in space where the displacement of the traveling wave fronts is still notable. It should be noted that due to this, the range of the parametric amplification waves is not only determined by the influence of the pumping field: Also the group velocity of the traveling waves plays a decisive role. Faster spin waves can achieve higher amplified propagation ranges due to the delay-dependence of the amplification.

The propagation range of a $20\,\mathrm{ns}$ long spin-wave packet interacting with such $200\,\mathrm{ns}$ long pumping pulse is shown in Fig.~\ref{Fig:Pamp-glob2}~(b) as a function of the applied pumping peak power. Hereby, the pumping power is given as the supercritical pumping power with respect to the instability threshold power $P_\mathrm{th,eff}$ in dB (spin-wave excitation frequency $f_\mathrm{s} = 6.5\,\mathrm{GHz}$, $\mu_0H_\mathrm{ext} = 20\,\mathrm{mT}$). As can be seen from the figure, for low supercriticalities, very large ranges exceeding $120\,\mu\mathrm{m}$ can be observed, exceeding more than 50 times the spin-wave wavelength of $\lambda \approx 2.3\,\mu\mathrm{m}$. With increasing pumping power the range decreases, due to the increasingly faster saturation of the amplifier by the faster parametric generation. The two data sets shown in Fig.~\ref{Fig:Pamp-glob2}~(b) differ in the relative timing between the spin-wave excitation and the beginning of the pumping pulse. For the first data set (black), the pumping starts at the same time than the spin-wave packet is launched whereas for the second data set (red), the spin waves are launched a few nanoseconds before the pumping starts. In the latter scenario, the spin waves can already travel along the waveguide and reach a finite, stationary amplitude at distances $y > 0$ before the pumping is applied\footnote{In terms of the delay dependence discussed in the local amplification, this initial intensity corresponds to the realization of a positive delay $\Delta t >0$ at a larger distance from the antenna when the pumping pulse is applied.}. Consequently, the propagation range of the waves is increased, as can be seen for a supercriticality of $3.5\,\mathrm{dB}$ by comparing the two series.

From the time-resolved measurements, it is also possible to evaluate the velocity with which the signal carried by the amplified wave front travels through the spin-wave waveguide and becomes detectable after reaching a certain threshold-intensity. For a measure of this, we define the effective signal velocity, which is given by the inverse of the time $\Delta t$ by which the rising edges of the amplified wave fronts visible in Fig.~\ref{Fig:Pamp-glob2}~(b) are displaced with respect to each other over a certain distance $\Delta y$ , i.e., $v_\mathrm{s} = \Delta y/\Delta t$. If a constant intensity is needed for detection, this velocity determines how long it takes until this intensity is reached at a given distance from the spin-wave source. This velocity is determined by the interplay of the spin-wave group velocity and the amplification rate, which determine how long it takes to build up the needed intensity at a given point in space. For our determination of this effective velocity, we extrapolate the time at which the spin-wave intensity has risen to $30\,\%$ of its saturation value from the rising edge, which is associated with the arrival of the amplified spin-wave packet at the 'detector', given by the BLS laser spot in this experiment. If this arrival time is plotted as a function of the distance to the antenna, the slope $\Delta t/\Delta y$ is the inverse  of the effective signal velocity defined above. The corresponding values extracted this way are shown in Fig.~\ref{Fig:Pamp-glob2}~(c) as a function of the supercritical pumping power. As can be seen, for a low supercriticality, i.e., where large ranges are achieved, the effective signal velocity is slow in comparison to the spin-wave group velocity of about $5.5\,\mu\mathrm{m}/\mathrm{s}$. For the first series, the signal velocity stays always below the group velocity and approaches it with increasing pumping power. As mentioned above, this is mediated by the fact that the amplified wave-front needs a certain time until it reaches the intensity level needed for detection. Due to the fact that the supercriticality is weak, the incremental change of the spin-wave intensity per unit time is small (cf. Eq.~\ref{Eq:Intvicth}) and it takes a comparable large time until the waves have been amplified to the detection limit. This leads to a seemingly slow 'propagation' of the detectable signal. In contrast, for the second series, where the spin-waves have been launched before the pumping starts, the observed signal velocity is always higher and can even exceed the spin-wave group velocity. This is a consequence of the stationary spin-wave intensity present in the waveguide due to the earlier launch of the spin-wave packet. For large supercriticalities the rise from this stationary level - which is not connected to the spin-wave group velocity itself - is very fast and, consequently, the velocity with which the amplified wave-front becomes detectable within the waveguide is very high. Thus, where the spin waves are already present with a weak amplitude below the detection limit but above the thermal level, their information 'propagates' trough the waveguide seemingly fast. It should be noted that this is not connected to a propagation of the wave packet faster than the group velocity itself - the detected waves do not need to travel, they are basically already present at the point of detection before the pumping starts and they are merely amplified to the detection limit. 

\begin{figure}[t!]
\center
{\caption{a) Time-resolved BLS intensity showing the temporal evolution of a parametrically amplified, externally excited spin-wave packet (black) together with the temporal evolution of the parametrically generated waves (red). The blue lines represent an analytical reconstruction of the measured intensity using the same rise and fall times (after \cite{Phd-Braecher}). The shading corresponds to the time the pumping pulse is applied. b) Schematic of a pulse scheme that allows for an efficient parametric amplification without the creation of a large noise due to parametric generation. During the shaded time-windows, the pumping field is applied.} 
 \label{Fig:Pamp-glob3}}
{\includegraphics[width=0.8\textwidth]{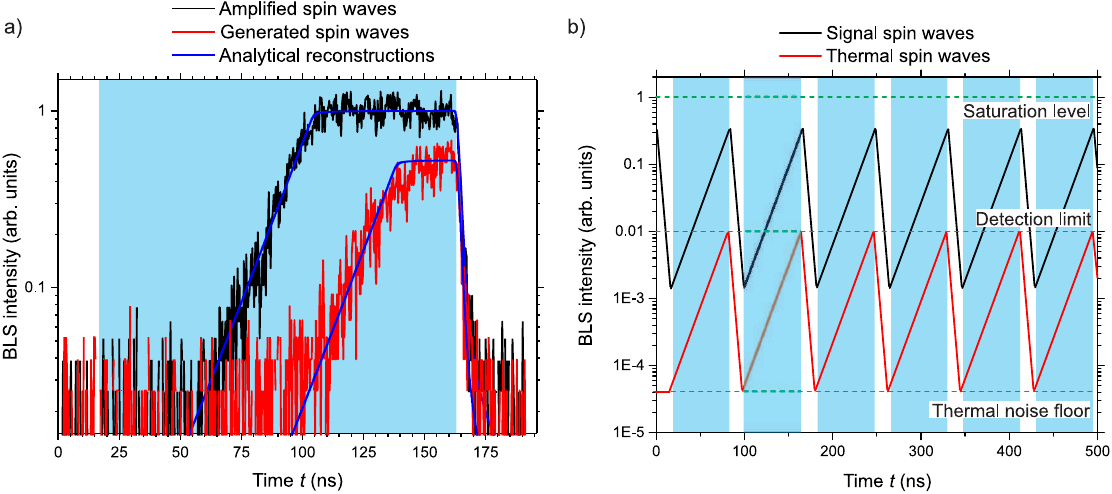}}
\end{figure}
Comparing Figs. \ref{Fig:Pamp-glob2}~(b) and (c), it becomes evident that - if a single pumping pulse is used for the spin-wave amplification - a trade-off between a long propagation range and a high effective signal velocity of the amplified waves has to be made. This way, propagation ranges beyond $70\,\mu\mathrm{m}$ with signal velocities exceeding $1\mathrm{km}\,\mathrm{s}^{-1}$ can be achieved. However, the fact that the signal and the dominant mode are the same opens an alternative path: Figure \ref{Fig:Pamp-glob3}~(a) shows the time-resolved spin-wave intensity of the amplified and the generated waves exemplarily\footnote{$f_\mathrm{sw} = 6.5\,\mathrm{GHz}$, $\mu_0 H_\mathrm{ext} = 20\,\mathrm{mT}$, $40\,\mathrm{ns}$ long spin-wave excitation pulses, $150\,\mathrm{ns}$ long pumping pulses.}. Due to the fact that the signal wave is the dominant mode, the two spin-wave packets feature identical rise and fall times. The solid blue lines in the figures are an analytical reconstruction of the measured spin-wave intensity taking into account the changes of the temporal profile due to the BLS detection and assuming the same rise and fall times for the amplified as well as the generated spin waves. As can be seen, they are both in good agreement with the measured profiles. The only difference in the temporal dynamics of these two spin waves is that the amplified spin-wave packet is detected at an earlier time because it starts with a larger starting amplitude than the thermal waves. Consequently, if a pulse scheme as sketched in Fig.~\ref{Fig:Pamp-glob3}~(b) is used, the parametric amplifier can be operated with a comparably large pumping power, resulting in a fast effective signal propagation as well as a large peak spin-wave intensity and, at the same time, it can operate with a very low noise. If the amplification pulses are short enough to avoid that the intensity of the parametrically generated waves always stays well below the initial level of the signal spin waves and if they are applied with a repetition that ensures that the externally excited, amplified packet always recovers to its initial amplitude during the amplification, a loss-less propagation without a notable creation of parametrically generated noise can be realized. As can be seen from Fig.~\ref{Fig:Pamp-glob3}~(b), where the rise and fall times of the experiment in Ni$_{81}$Fe$_{19}$ have been assumed, this scheme would require a high repetition rate of the pumping pulses. Even though a fast enough on- and off-switching of the pumping might be feasible by a temporal modulation of the pumping power, for a practical application, a material with a lower spin-wave damping is favorable. Such materials, like the half-metallic Heusler compound CoMnFeSi\cite{Sebastian-2012} or microstructured YIG\cite{Pirro-2014-1,Hahn-2014-1} will allow for a smaller repetition rate and a larger tolerance of the pulses. 

\section{Novel concepts and applications}
\label{Nov}

In the previous chapter, the prospects and the limitations of the application of parallel pumping as an exclusive means of spin-wave amplification have been discussed. In the following Chapter, some of its applications that go beyond a mere amplification will be presented. 

\subsection{Phase-to-intensity conversion}
\label{Nov-PtI}

From Eq.~\ref{Eq:Phaserel} it follows that the phase of the two created spin waves, the signal as well as the idler spin waves, are mutually connected to the phase of the pumping field. Since the phase of the pumping field and the phase of the signal spin waves are externally determined quantities, the idler wave needs to adjust its phase in order to fulfill this phase-relation. If the phase evolution between the signal spin waves and the parallel pumping field is externally locked, this can be used to convert the phase of the signal spin waves into an intensity information. In a global amplifier such as the one discussed above, this implies that the standing wave pattern arsing from the interference of the counter-propagating signal and idler waves can be shifted in space by changing the relative phase $\Delta \phi_\mathrm{sp} = 2\phi_\mathrm{s} - \phi_\mathrm{p}$ between the signal spin waves and the pumping field. Consequently, the phase of the idler wave has to adjust and, hence, the relative phase $\Delta \phi_\mathrm{si} = \phi_\mathrm{s} - \phi_\mathrm{i}$ between signal and idler waves changes, which displaces the standing wave pattern in space. Therefore, within the amplifier, the intensity of the standing wave is connected to the relative phase between the signal spin waves and the pumping field. Thus, in the case of the creation of counter-propagating waves, which is commonly referred to as adiabatic parametric amplification\cite{Melkov-Parametric-interaction,Melkov-nonadiabatic}, the use of a detector in a fixed position allows to deduce on the phase of the signal spin waves if the phase of the pumping field is employed as a fixed reference phase.

\begin{figure}[h]
\center
{\caption{Working principle and concept of the phase-to-intensity conversion by nonadiabatic parallel pumping. a) Due to the finite size of the amplifier, it provides a spectrum of effective wavevectors $k_\mathrm{p}$ (green line). b) Microwave photons with $k_\mathrm{p}>0$ can split into co-propagating magnon pairs, leading to the formation of the signal and idler waves at one half of the pumping frequency. The solid dark blue line represents the spin-wave dispersion relation of the fundamental waveguide mode in the investigated Ni$_{81}$Fe$_{19}$ waveguide. c) For a pumping field with fixed reference phase, schematically illustrated by the green line, the signal (violet) and idler (orange) waves are initially in-phase (solid lines). In this case, their interference is constructive. If the phase of the signal wave is shifted by $\pi/2$, the phase of the idler wave has to adjust its phase by $-\pi/2$ (dashed lines). Consequently, their interference becomes destructive. d) Output intensity of the amplifier resulting from this interference mechanism (after \cite{PhasetoInt}).} 
 \label{Fig:Nov-PtI1}}
{\includegraphics[width=1.0\textwidth]{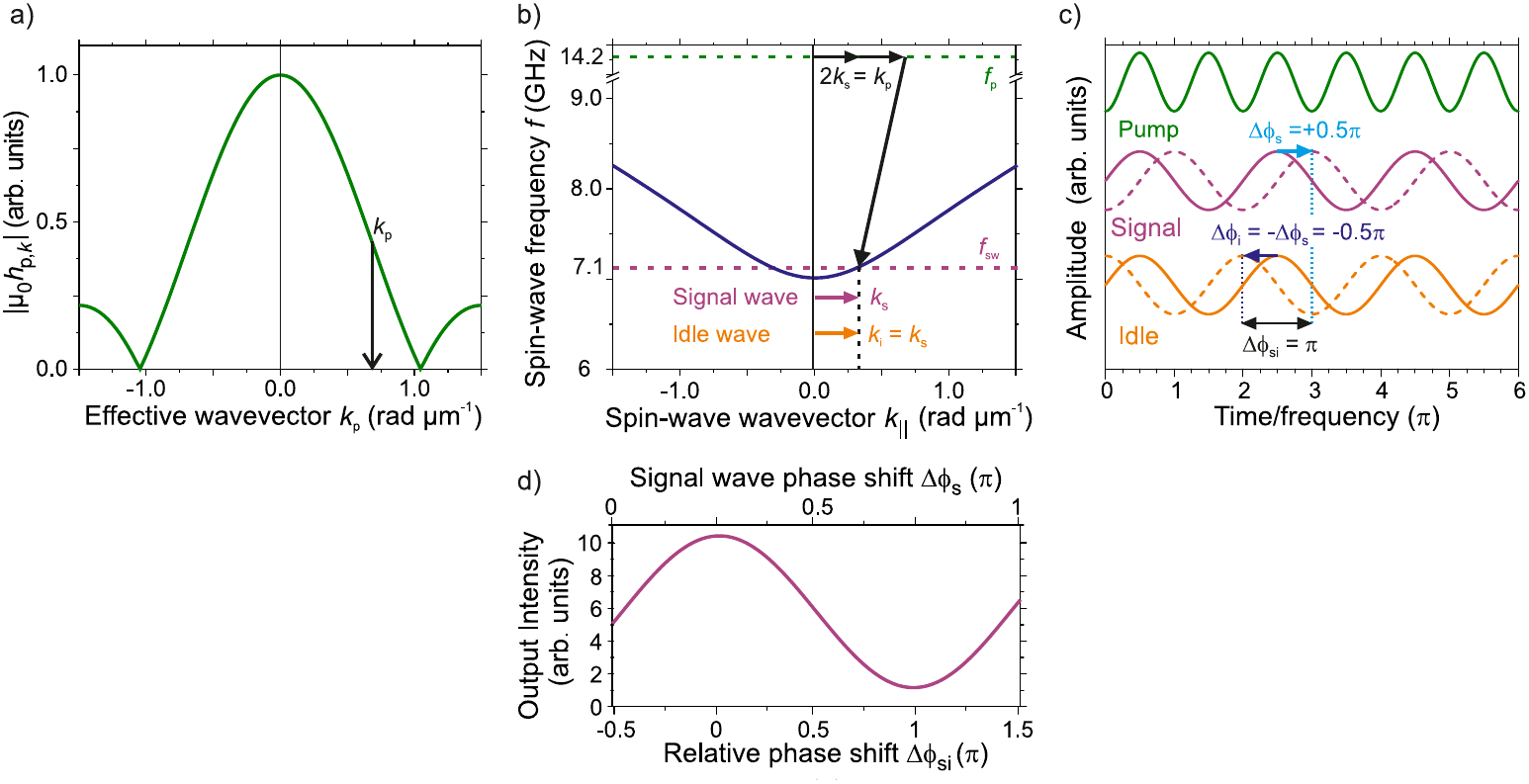}}
\end{figure}

Within the amplifier it is possible to connect the phase of the signal spin waves to an intensity at a given position. If the amplifier is not extending along the entire magnetic structure but is still large in comparison to the spin-wave wavelength, the output of the amplifier in the forward direction is, however, not affected by the relative phase between signal spin waves and pumping field. Hence, it is impossible to obtain a phase information which is not sensitive to the readout position or to pass on only spin waves with a desired phase within a magnonic network. In turn, this can be achieved by making the amplifier smaller than the wavelength of the spin waves which are to be amplified. In this \textit{nonadiabatic regime} of parametric amplification, the amplification process is quite different. The basic features of this regime are shown in Fig.~\ref{Fig:Nov-PtI1}.

Figure \ref{Fig:Nov-PtI1}~(a) shows the Fourier spectrum of a $6\,\mu\mathrm{m}$ long rectangular amplifier in wavevector space. As can be seen from the figure, the spatial confinement leads to the presence of an extensive effective wavevector spectrum of the parametric amplifier. These finite wavevectors allow for the splitting process schematically shown in Fig.~\ref{Fig:Nov-PtI1}~(b): A photon with an effective wavevector of $k_\mathrm{p} = 2k_\mathrm{s}$ splits into two co-propagating signal and idler waves with wavevector $k_\mathrm{s}$. Hereby, the finite effective wavevector $k_p$ does not imply that the pumping field consists of a propagating wave. This just accounts for the fact that, throughout the amplifier, the pumping field can provide the necessary momentum to fulfill the momentum conservation condition stated in Eq.~\ref{Eq:conservation-nonadiabatic} at all times. Thus, in this nonadiabatic regime, the creation of co-propagating wave pairs is possible. The consequence of this wave-pair formation is shown in Fig.~\ref{Fig:Nov-PtI1}~(c). If the phase of the pumping field is considered as reference and the signal and the idler waves are initially in phase ($\Delta \phi_\mathrm{si} = 0$), this leads to a large output of the amplifier. However, if the phase of the signal wave is shifted by $\pi/2$, the idler wave has to adjust its phase by $-\pi/2$ in order to fulfill Eq.~\ref{Eq:Phaserel}. Consequently, the created spin waves are out-of-phase and the output of the amplifier is minimal ($\Delta \phi_\mathrm{si} = \pi$). Fig.~\ref{Fig:Nov-PtI1}~(d) shows the time-averaged output intensity of the amplifier arising from this interference mechanism which, in the absence of the creation of noise due to parametric generation, is given by the interference of the incoming signal spin wave with the signal and idler spin waves created by the parallel pumping process. For amplitudes $A_\mathrm{s} (t)$ of the signal wave and $A_\mathrm{i}(t)$ of the idler wave, the time-averaged output intensity of the amplifier at a fixed readout position is given by:
\begin{align}
\begin{split}
\label{Eq:Output}
I &= \overline{|A_\mathrm{s}(t) + A_\mathrm{i}(t)|^2}^\mathrm{t} \\
  &= A_0 + (A_\mathrm{s,0} + A_\mathrm{a})\cdot A_\mathrm{a}\cdot \cos(\Delta \phi_\mathrm{si})\\
	&= A_0 + (A_\mathrm{s,0} + A_\mathrm{a})\cdot A_\mathrm{a}\cdot \cos(\Delta \phi_{\mathrm{sp}}-\pi/2),
\end{split}
\end{align}
with the constant $A_0 = ((A_\mathrm{s,0} + A_\mathrm{a})^2+A_\mathrm{a}^2)/2$\footnote{In the presence of a strong amplification, an additional amplification of counter-propagating waves or parametric generation can constitute an additional offset in the equation.}. Here, $A_\mathrm{s,0}$ denotes the amplitude of the incoming signal wave and $A_\mathrm{a}$ the amplitude of the created signal as well as idler waves. Hence, the output of the amplifier can be tuned continuously by changing the relative phase $\Delta \phi_\mathrm{sp}$. It should be noted that the dependence of the output intensity on $\Delta \phi_\mathrm{sp} = 2\phi_\mathrm{s} - \phi_\mathrm{p}$ implies that the output of the amplifier is invariant to a change of the spin-wave phase of $\pi$. Consequently, the spin-wave phase has to be encoded in values between 0 and $\pi/2$ for the application of such an amplifier to a logic architecture relying on the phase of the spin waves as continuous variable\cite{PostCMOS3,Majority1,Maj,Klingler_majority-gate1,Klingler_majority-gate2}.

For the experimental realization of a phase-dependent amplification in microstructured spin-wave waveguides, a local spin-wave amplifier such as the one schematically shown in Fig.~\ref{Fig:Pamp-Loc1} can be used. If the spin-wave excitation and the parallel pumping are phase-locked, it is possible to observe the influence of the relative phase between the signal spin waves and the pumping field $\Delta \phi_\mathrm{sp}$ on the output of the amplifier\cite{Melkov-Parametric-interaction,Melkov-nonadiabatic} if the spin-wave wavelength is comparable to or larger than the size of the local amplifier.

\begin{figure}[h]
\center
{\caption{a) BLS intensity (logarithmic scale) as a function of the distance to the micro-strip antenna. Green rectangles and blue upward triangles: Spin-wave intensity arising from the interplay between the nonadiabatic parametric amplification and the spin-wave excitation. Between these two measurements, the phase of the pumping field is shifted. Red circles: Pumping only. Violet downward triangles: spin-wave excitation at the antenna only. The shaded area represents the spatial extent of the amplifier. b) Intensity measured at a distance of $7\,\mu\mathrm{m}$ (marked by the arrow in (a)) behind the center of the nonadiabatic amplifier as a function of the induced phase shift $\Delta \phi_\mathrm{sp}$ between the signal wave and the pumping field as well as of the resulting phase shift $\Delta \phi_\mathrm{si}$ between the signal and the idler waves. The violet line represents a fit according to Eq.~\ref{Eq:Output} (after \cite{PhasetoInt}). c) Intensity along the waveguide in the case of adiabatic parametric amplification. In between the two measurements, the relative phase between the spin-wave excitation and the pumping field is shifted.} 
 \label{Fig:Nov-PtI2}}
{\includegraphics[width=1.0\textwidth]{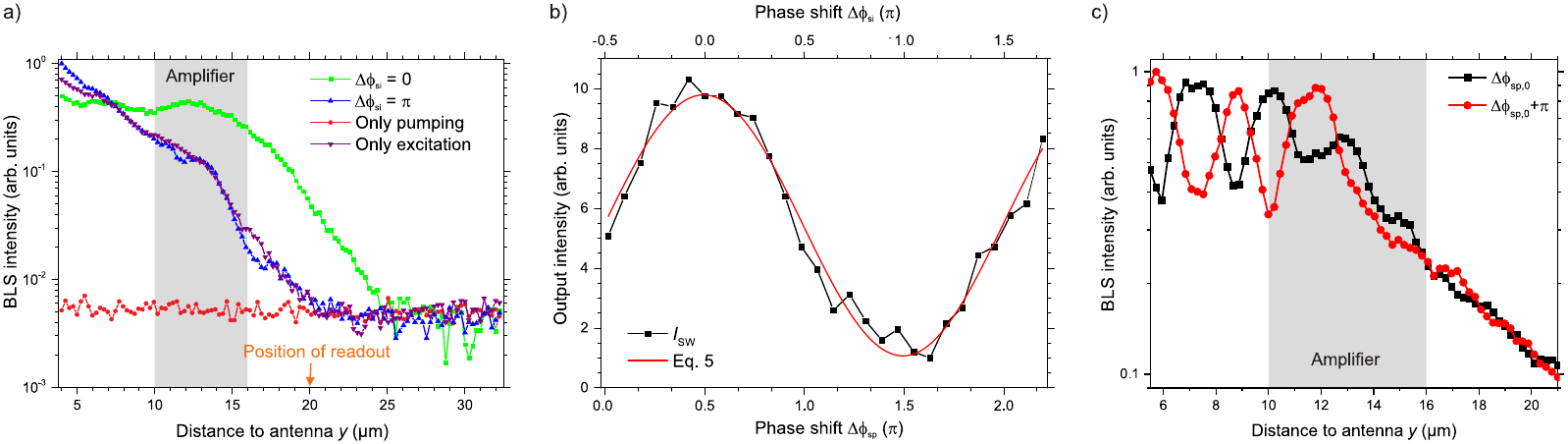}}
\end{figure}

Figure \ref{Fig:Nov-PtI2}~(a) shows the spin-wave intensity measured along the center-line of a $4.2\,\mu\mathrm{m}$ wide, $40\,\mathrm{nm}$ thick, transversely in-plane magnetized Ni$_{81}$Fe$_{19}$ waveguide for different scenarios. In this experiment, which is described in more detail in Ref. \cite{PhasetoInt}, the signal spin waves are excited with a $50\,\mathrm{ns}$ long microwave pulse with a carrier frequency of $7.13\,\mathrm{GHz}$ and their interaction with $50\,\mathrm{ns}$ long pumping pulses with twice this frequency is studied. The pumping pulse starts when the spin-wave packet enters the amplifier, and the used peak pumping power of $P = 22\,\mathrm{dBm}$ is sufficiently low to avoid a strong parametric generation during the duration of the pumping pulse. At the applied external bias field of $\mu_0H_\mathrm{ext} = 63\,\mathrm{mT}$, the spin-wave frequency corresponds to a spin-wave wavelength of $\lambda = 13.5\,\mu\mathrm{m}$, which is larger than the local parametric amplifier created by a $6\,\mu\mathrm{m}$ long and $4\,\mu\mathrm{m}$ wide geometric constriction. Consequently, the output of the amplifier will depend on the relative phase between the signal spin wave and the pumping field. 

The green squares show the spin-wave intensity which arises from the parametric amplification of the input spin waves for a constructive interference between the signal and the idler spin waves ($\Delta \phi_\mathrm{si} = 0$, which corresponds to $\Delta \phi_\mathrm{sp} = 0.5\,\pi$). As can be seen from Fig.~\ref{Fig:Nov-PtI2}~(a), in this scenario, the spin waves are amplified during their propagation through the amplifier, similar to the adiabatic case shown in Fig.~\ref{Fig:Pamp-Loc1}~(b). If the relative phase of the signal spin waves is shifted by $\pi/2$ with respect to the phase of the pumping field via a phase-shifter localized in the microwave setup providing the pumping pulses, the interference between the signal and the idler waves becomes destructive ($\Delta \phi_\mathrm{si} = \pi$, $\Delta \phi_\mathrm{sp} = 1.5\pi$). The resulting spin-wave intensity is shown by the blue upward triangles. In contrast to the amplification with a constructive interference, a much smaller spin-wave intensity is leaving the amplifier and, in agreement with Eq.~\ref{Eq:Output}, the intensity within and after the amplifier basically coincides with the intensity arising from the unamplified spin-wave packet, which is shown by the violet downward triangles. For comparison the red dots show the BLS intensity if only the pumping field is applied. Due to the low pumping power in combination with the finite duration of the pumping pulse, no parametrically generated noise is visible, showing that such a parametric amplifier can work with a low operational noise. 

Figure~\ref{Fig:Nov-PtI2}~(b) shows the output of the amplifier as a function of the relative phase between the signal spin waves and the pumping field together with a cosinusoidal fit according to Eq.~\ref{Eq:Output}. The output is measured at a distance of $7\,\mu\mathrm{m}$ from the center of the amplifier. As can be seen from the figure, the output can be continuously tuned as a function of the phase shift. Thus, if the phase of the pumping field is kept fixed with a certain reference value, the phase of the signal spin waves is directly transferred into an amplified intensity. This intensity can be readout conveniently by conventional detection schemes such as the inverse Spin Hall effect in combination with spin pumping or the tunneling magnetoresistance effect\cite{Spin-Pumping,iSHE,Sandweg,iSHE2,iSHE3,MTJ}, which convert the AC spin-wave intensity into a DC voltage. Hereby, the inherent amplification of the spin waves with proper phase facilitates the readout and allows to operate with a low spin-wave intensity in the conduit, minimizing energy costs and nonlinear spin-wave interaction. It should be noted that only an amplitude of amplified signal and idler waves $A_\mathrm{a}$ comparable to the amplitude of the incident spin waves $A_0$ is needed to achieve such a large modulation ($A_\mathrm{a} \approx 1.1 A_0$ in the figure). If the pumping power is increased, a lager number of amplified waves $A_\mathrm{a}$ can easily be achieved (i.e., high saturation level). To minimize the noise and to maximize the intensity difference between constructive and destructive interference, a proper pulse duration to reduce the effect of the stronger parametric generation at larger pumping powers should be used (cf. discussion at the end of Section \ref{Pamp-glob}).

For comparison, Fig.~\ref{Fig:Nov-PtI2}~(c) shows the BLS intensity measured along the spin-wave waveguide if the magnetic bias field is changed to $\mu_0H_\mathrm{ext} = 50\,\mathrm{mT}$ for two different phase-relations between the signal waves and the pumping field. For this bias field, the spin-wave wavelength is reduced to $\lambda = 5.1\,\mu\mathrm{m}$ and, consequently, the parametric amplifier operates in the adiabatic regime. As mentioned above, in this regime, a change of the relative phase should only change the interference pattern within or in front of the amplifier and should not affect the output of the amplifier. As can be seen from the two measurements shown in Fig.~\ref{Fig:Nov-PtI2}~(c), between which the relative phase between the signal spin waves and the pumping field has been shifted by $\Delta \phi_\mathrm{sp} = \pi$, this behavior is well reproduced in the experiment. Thus, with an operation in the adiabatic regime, spin waves are amplified in the forward direction independently of their phase, while a readout in front of the amplifier allows for a deduction on their phase based on the detected spin-wave intensity. Due to the fact that the transition from the adiabatic to the nonadiabatic operation regime can be achieved by simply changing the spin-wave wavelength, a local parametric amplifier is, thus, a very versatile tool in any magnonic network. 

\begin{figure}[t] 
\center
{\caption{a) Schematic of the simulated majority gate featuring a parametric amplifier to facilitate the readout of the phase information. The wavy arrows represent input spin waves with $\Delta \phi_\mathrm{s,0}= 0$ (logic 0, green) and $\Delta \phi_\mathrm{s,0}=0.5\pi$ (logic 1, orange), respectively. b) Output spin-wave intensity as a function of the offset of the phase of the signal spin wave $\Delta \phi_\mathrm{s,0}$ in the output waveguide. The two curves are simulated for different phase offsets $\Delta \phi_\mathrm{p,0}$ of the pumping relative to the spin-wave excitation, favoring either the amplification of logic 0 or logic 1. The solid lines represent Eq.~\ref{Eq:Output} (after \cite{PhasetoInt}).} 
 \label{Fig:Nov-PtI3}}
{\includegraphics[width=0.8\textwidth]{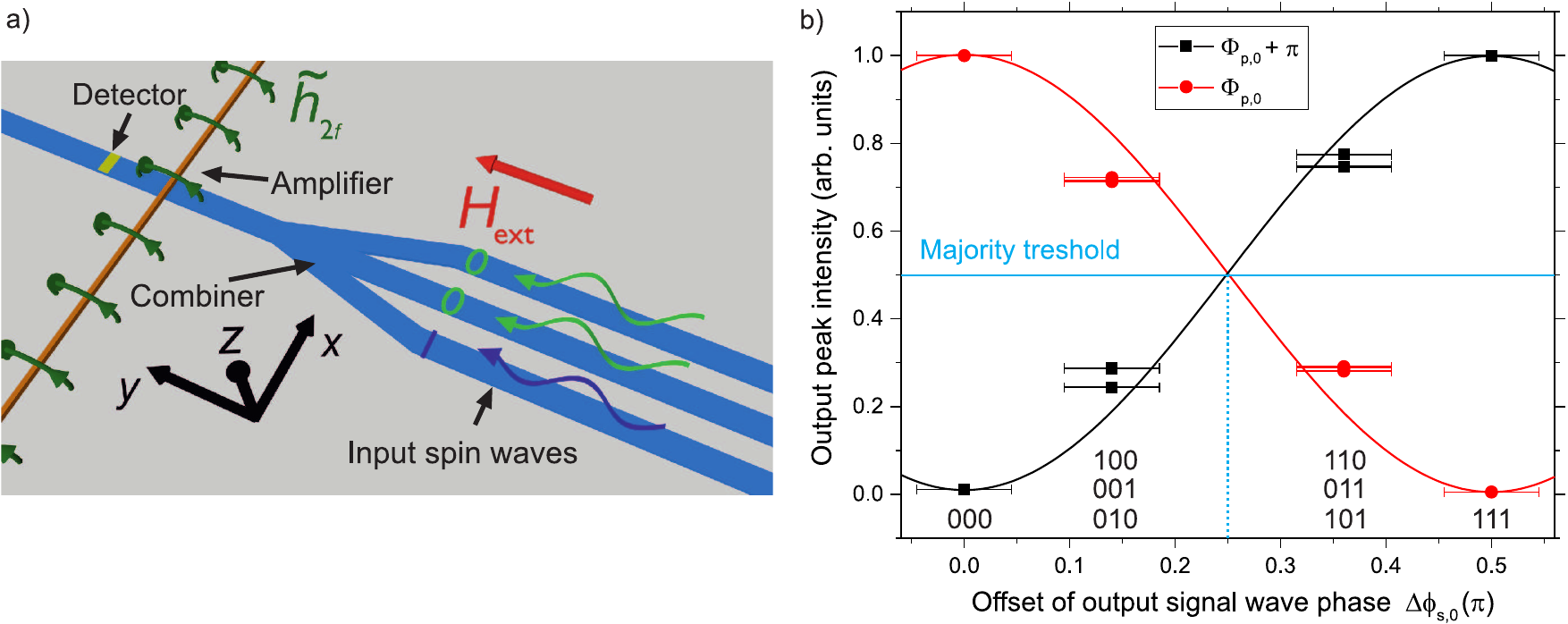}}
\end{figure}
One instructive example of how a nonadiabatic parametric amplifier can be implemented into a magnonic network has been demonstrated in Ref. \cite{PhasetoInt}. In this work, the application of the phase-to-intensity conversion by parallel pumping to a magnonic majority gate\cite{Maj,Klingler_majority-gate1,Klingler_majority-gate2} has been studied numerically by micromagnetic simulations using MuMax3\cite{MuMax-2014-1}. A majority gate\cite{PostCMOS3,Majority1} is a logic element with an odd number of inputs whose output represents the majority of the input values. It can act as the basic element of novel logic architectures. Figure~\ref{Fig:Nov-PtI3}~(a) shows the layout of the simulated device: Three input waveguides are merged into an output waveguide in a combiner region (see also Ref. \cite{Klingler_majority-gate1}). All waveguides operate in the backward-volume geometry and are longitudinally magnetized by an external field of $\mu_0 H_\mathrm{ext} = 50\,\mathrm{mT}$. Each input can be connected to another magnonic device or a spin-wave excitation source. In the output of the majority gate, the dynamic field of a $200\,\mathrm{nm}$ wide micro-strip antenna is simulated, which acts as local, nonadiabatic source of the pumping field (cf. Section \ref{Pgen-BV}). If a logic zero is encoded in a spin-wave phase-shift of $0$ and a logic one is encoded in a phase-shift of $0.5\pi$, the majority operation is preserved, if all phase-shifts $>0.25\pi$ are considered as majority 1. The output of the majority gate with parametric amplification is shown in Fig.~\ref{Fig:Nov-PtI3}~(b), together with a fit according to Eq.~\ref{Eq:Output}. As can be seen from the figure, for a fixed phase of the pumping field, the pumping results in four distinct intensity levels, representing the different combinations of the input waveguides. By defining the majority by a threshold intensity situated at one-half of the maximum output intensity, the parametric amplifier preserves the majority function. Moreover, from the absolute value of the intensity it can be concluded whether the majority is full (e.g., input 1 in all input waveguides, abbreviated as 111) or partial (e.g., 101, i.e., 1 in input 1 and 3 and 0 in input 2). Moreover, if the phase of the pumping field is shifted by $\pi$, the desired majority which should be amplified can be selected, allowing for a flexible operation of the majority gate. This is just one illustrating example on how the use of the technique of nonadiabatic parallel pumping can enhance the functionality of a magnonic circuit.

\subsection{Determination of the spin-wave dispersion relation and the phenomenon of nonreciprocity in individual microstructures}
\label{Nov-Disp}

A caveat of micro-focused BLS is the loss of direct information about the wavevector of the investigated spin waves. This information can be recovered by phase-resolved BLS\cite{Schneider-PR, Vogt-PR}, i.e., the interference of the inelastically scattered photons with photons with a well-defined frequency modulation, or by investigating the interference of counter-propagating spin waves\cite{Pirro-2011-1}. Parametric amplification provides an interesting alternative to these techniques, which does not require the use of an electro-optical modulator, which limits the usable frequency-range significantly, or the use of a second antenna. The inherent interference of the counter-propagating waves in an adiabatic amplifier provides the wavevector information with some additional advantages: On the one hand, the parametric amplification enhances the propagation range of the waves remarkably, which gives access to the observation of the interference of longer wavelengths closer to $k \approx 0$. On the other hand, due to its weak wavevector selectivity, the process of parallel pumping can amplify even a very weakly excited spin-wave intensity beyond the efficient excitation range of a micro-strip antenna and, this way, gives experimental access to a larger wavevector span. 

\begin{figure}[h]
\center
{\caption{a) Spin-wave frequency as a function of the spin-wave wavevector determined from interference measurements for a fixed bias field $\mu_0H_\mathrm{ext} = 40\,\mathrm{mT}$. The red line is calculated according to Eq.~\ref{Eq:Disp-tf}. b) Spin-wave wavevector as a function of the applied bias field for a fixed spin-wave frequency of $f = 6.5\,\mathrm{GHz}$. The  red line is calculated according to Eq.~\ref{Eq:Disp-tf} and the blue line is a linear fit to the data. c) Adiabatic parametric amplification and a non-reciprocal dispersion in transversely magnetized waveguide: To conserve momentum in the pumping process, spin-wave pairs with a frequency splitting are created. d) Time-resolved BLS intensity showing the intensity modulation arising from the interplay of a $500\,\mathrm{ns}$ long spin-wave packet with the pumping field. The red line is a cosinusoidal fit (see text). e) Frequency splitting as a function of the spin-wave wavevector determined for different frequency-and-field combinations. The orange line constitutes a theoretical modeling following Ref. \cite{Burkard-PRL-1990} (after \cite{Phd-Braecher}).} 
 \label{Fig:Nov-Disp}}
{\includegraphics[width=1.0\textwidth]{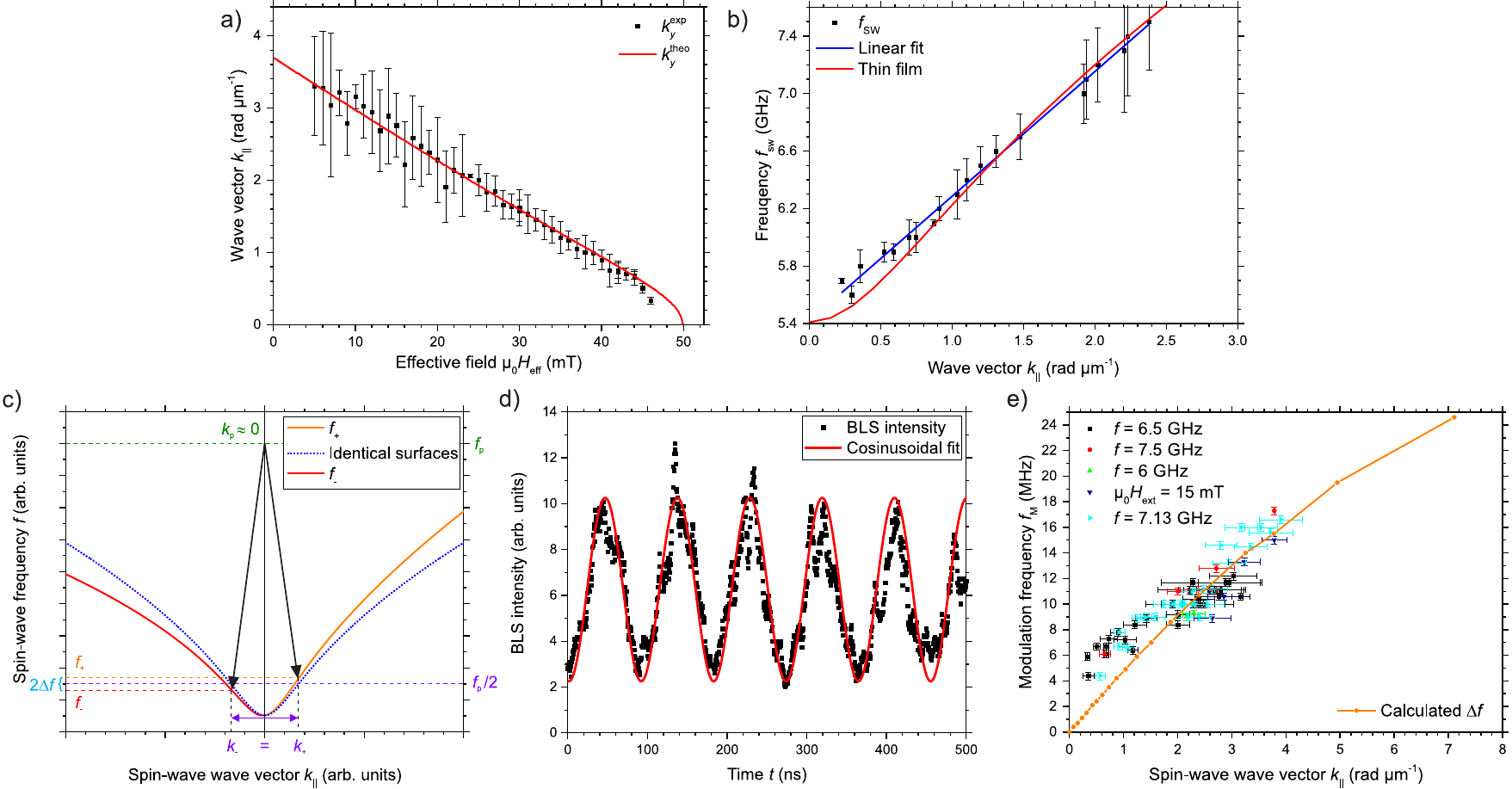}}
\end{figure}

Figs.~\ref{Fig:Nov-Disp}~(a) and (b) show the spin-wave wavevector obtained from interference patterns such as the one shown in Fig.~\ref{Fig:Pamp-glob1}~(b) as a function of the applied external bias-field for a fixed frequency of $f = 6.5\,\mathrm{GHz}$ (a) and as a function of the applied frequency for a fixed field of $\mu_0H_\mathrm{ext} = 40\,\mathrm{mT}$ (b) together with the analytically calculated dispersion relations of the fundamental mode according to Eq.~\ref{Eq:Disp-tf}. The measurements have been performed on a $4.2\,\mu\mathrm{m}$ wide, $30\,\mathrm{nm}$ thick magnonic waveguide made from Ni$_{81}$Fe$_{19}$ grown on a thermally oxidized Si wafer on top of a Ru/MgO buffer layer and the waveguide has been capped with MgO. The waveguide has been magnetized transversely by an external bias field and the signal spin waves have been excited with a $2.2\,\mu\mathrm{m}$ wide antenna, which is able to excite spin waves efficiently in a range with $k_\mathrm{||, max} \lesssim 2.9\,\mathrm{rad}\,\mu\mathrm{m}^{-1}$. The pumping field and the spin-wave excitation have been performed by $20\,\mathrm{ns}$ long microwave pulses and the microwaves have been created by the same frequency generator together with a frequency doubler for the pumping field to ensure a fixed phase-relation (cf. Section \ref{Pamp-glob}). The analytical calculations have been performed using $M_\mathrm{s} = 830\,\mathrm{kA}\,\mathrm{m}^{-1}$, $A_\mathrm{ex} = 13\,\mathrm{pJ}\,\mathrm{m}^{-1}$, $w_\mathrm{eff} = 3.6\,\mu\mathrm{m}$ and $\mu_0H_\mathrm{Demag} = 5\,\mathrm{mT}$ ($\mu_0H_\mathrm{eff} = \mu_0H_\mathrm{ext}-\mu_0H_\mathrm{Demag}$). As can be seen from the figure, the analytical calculations provide a reasonable agreement for wavevectors $k_{||}\gtrsim 1\,\mathrm{rad}\,\mu\mathrm{m}^{-1}$, whereas they provide a poor description for smaller wavevectors. In fact, the frequency vs. wavevector dependence can be approximated by a linear dependence, a finding that has also been made in Ref. \cite{Chang2014}. This is related to the fact that Eq.~\ref{Eq:Disp-tf} is only an approximation, which, among others, neglects the dipolar interaction between the waveguide modes. Since, in addition, the use of Eq.~\ref{Eq:Disp-tf} requires the knowledge of a large set of material parameters, the convenient characterization of the spin-wave dispersion by the adiabatic parallel pumping process provides a useful tool for experimentalists.

The Ni$_{81}$Fe$_{19}$ waveguide investigated in the study of the spin-wave dispersion depicted in Figs. \ref{Fig:Nov-Disp}~(a) and (b) features growth-induced differences between the bottom and the top interface\cite{Gladii-2016}. This gives rise to different surface anisotropy-constants at the individual surfaces which, in turn, leads to a splitting of the spin-wave dispersion for spin waves running with $\textbf{k} \perp \textbf{M}$ along the waveguide\cite{Burkard-PRL-1990, Gladii-2016}. However, the observed frequency-splitting is on the order of a few MHz and, thus, way below the frequency resolution of the used BLS spectrometer. One way to work around this limitation in the frequency resolution is the use of parametric amplification of coherently excited traveling spin waves together with micro-focused BLS or electrical detection. The concept of the detection of such small frequency-non-reciprocities in transversely magnetized waveguides is schematically shown in Fig.~\ref{Fig:Nov-Disp}~(c): The presence of different interfaces leads to different boundary conditions for waves traveling to the left and to the right along the waveguide. With respect to the dispersion relation $f_0(k_{||})$ of the fundamental mode in the case of identical interfaces, which is indicated by the dashed lines, this leads to a splitting of the dispersion for $\pm k_\mathrm{||}$ which is an odd function of the spin-wave wavevector\cite{Gladii-2016}. In the process of adiabatic parallel pumping, the two created waves need to satisfy $\mathbf{k}_+ + \mathbf{k}_- = 0$. Here, $'+'$ denotes the wave traveling to the right and $'-'$ the wave traveling to the left. If the frequency-splitting is an odd function of $k_\mathrm{||}$, the system satisfies momentum and energy conservation by amplifying wave pairs featuring $|\mathbf{k}_+| = |\mathbf{k}_-| = k$ and $f_+ =f_0+\Delta f(k)$ and $f_- =f_0-\Delta f(k)$ with $f_0 = f_\mathrm{p}/2$, which is possible in many situations due to approximate linearity of the spin-wave dispersion relation (cf. Fig.~\ref{Fig:Nov-Disp}~(a) and Ref. \cite{Chang2014}). 

Hence, the interference pattern resulting from the interference of the counter-propagating signal and the idler waves will be non-stationary, since now the difference between the two frequencies does not vanish. Instead, the created interference pattern, which features a spatial periodicity of $2k$, is now constantly moving in space with the difference of the frequencies $f_+-f_- = 2\Delta f$  since $I(y,t) \propto \cos(2ky + 2 \Delta f t + \phi_0)$. Thus, by measuring the time-resolved BLS intensity in a fixed position, one observes a temporal modulation of the BLS intensity with a frequency in the MHz range. This modulation can easily be resolved with the temporal resolution of BLS\footnote{If the frequency splitting becomes so large that the modulation is faster than the time-resolution, it is, in turn, again visible in the frequency domain and will manifest as satellites of the main-peak in the BLS spectrum with $\pm \Delta f$.}, which lies in the order of $1\,\mathrm{ns}$. 

This is demonstrated in Fig.~\ref{Fig:Nov-Disp}~(d), where the time-resolved BLS intensity shows the interplay of a $500\,\mathrm{ns}$ long spin-wave packet ($\mu_0H_\mathrm{ext} = 20\,\mathrm{mT}$, $f = 6.5\,\mathrm{GHz}$, $\lambda = 2.3\pm0.4\,\mu\mathrm{m}$) with the microwave pumping field together with a cosinusoidal fit. It should be noted that the intensity modulation can be maintained on these comparably long time scales only because the dominantly generated spin-wave mode in the waveguide is the fundamental mode. The fit yields $\Delta f = (11.1\pm0.1)\mathrm{MHz}$, a value in the order of magnitude expected for a Ni$_{81}$Fe$_{19}$ film in this thickness range with adjacent oxide interfaces\cite{Gladii-2016}. Figure \ref{Fig:Nov-Disp}~(e) shows the wave-vector dependent frequency splitting obtained from similar measurements performed as a function of the externally applied bias field for several fixed frequencies and as a function of frequency for a fixed field of $\mu_0 H_\mathrm{ext} = 15\,\mathrm{mT}$. For all these measurements, the wavevector $k_\mathrm{||}$ has been extracted from the spatial modulation from measurements with short pulses. The frequency shifts have been determined from the Fourier transformation of the time-resolved BLS measurements. As can be seen from the figure, the non-reciprocity primarily depends on the wavevector $k_{||}$, as the values for $\Delta f(k)$ obtained from the different measurements mostly coincide within the error bars. The orange curve is a theoretical prediction of the frequency non-reciprocity based on Ref. \cite{Burkard-PRL-1990} assuming nonidentical surface constants of $K_{\perp,1} = 0.6\,\mathrm{mJ}\,\mathrm{m}^{-2}$ and $K_{\perp,2} = 0.45\,\mathrm{mJ}\,\mathrm{m}^{-2}$, values which are in the expected parameter range \cite{Ruiz-2015,Gladii-2016}. As can be seen from the figure, the model provides a reasonable description for wavevectors larger than $k \approx 2\,\mathrm{rad}\,\mu\mathrm{m}^{-1}$, whereas there is a significant deviation for smaller wavevectors. Thus, the region at very small wavevectors, which is not accessible to conventional BLS measurements, indicates that a more involved theoretical model is necessary to explain all the (nonreciprocal) features of the spin-wave dispersion in such layer systems.

Another origin of frequency non-reciprocities which are an odd function of the spin-wave wavevector is the interfacial Dzyaloshinskii-Moriya interaction (iDMI)\cite{Di-2015, Moon-2013, Kostylev-2014, Nembach-2015,Stashkevich-2015}, which can arise in thin magnetic films in a layer system featuring a structural inversion asymmetry (SIA)\cite{Fert-1990,Fert-2013}. The iDMI has recently attracted a large research interest due to its intriguing consequences such as, among others, the stabilization of magnetic skyrmions\cite{Fert-2013,Bergmann-2014,Rohart-2013,Boulle-2016}, its impact on the domain wall dynamics \cite{Boulle-2013,Martinez-2013,Brataas-2013,Emori-2013,Ryu-2013} or the deterministic magnetization switching driven by spin orbit torques\cite{Nikolai-2015}.

Conventional measurements of the iDMI-strength in ultra-thin films are either performed by analyzing the domain-wall expansion\cite{Hrabec-2014-1,Je2013} or by analyzing the non-reciprocal spin-wave dispersion by conventional, wavevector resolved Brillouin light scattering spectroscopy in extended films\cite{Nembach-2015,Stashkevich-2015,Di-2015}. More recently, also an electrical characterization of the iDMI in an individual microstructure has been reported in Ref. \cite{Lee-2016} using similar means as have been used in Ref. \cite{Gladii-2016} for the characterization of influence of unequal surface anisotropies. While the analysis of the domain-wall-motion in the presence of iDMI can be rather involved, the BLS measurements suffer from the rather small frequency resolution of the BLS of a few $100\,\mathrm{MHz}$, which, as mentioned above, hampers the determination of the iDMI-constant (or anisotropy-differences) for small wavevectors or for very small iDMI-strengths. Moreover, since micro-focused BLS comes along with a loss of wavevector resolution, this method cannot be used to probe individual micro- or nanostructures unless phase-resolved BLS is applied or spin-wave interference is studied. The demonstrated electrical detection schemes can be applied to individual microstructures. However, the demonstration in Ref. \cite{Lee-2016} has been carried out on Pt/Co/MgO, a system that features a very high DMI constant, and the method is already at the detection limit in this system. Since the impact of the iDMI on the spin-wave dispersion scales with the inverse of the thickness of the ferromagnetic layer, its determination is best performed in very thin films, which also helps to reduce the influence of the likely unequal surface anisotropies in the presence of SIA. Due to the large induced damping caused by the contact to the heavy metals in common systems featuring iDMI such as Co/Pt, Ni$_{81}$Fe$_{19}$/Pt or CoFeB/Pt and the low group velocity in very thin films, propagating spin-wave spectroscopy becomes a challenging task. Here, the presented method to use the interference pattern created by the parametric amplification of traveling waves offers a good alternative: It can detect very tiny shifts and it is not limited by the small propagation range. Thus, together with the determination of the spin-wave dispersion using short microwave pulses to determine the spin-wave wavevector from the spatial interference pattern, it constitutes a valuable alternative method for the characterization of iDMI in individual microstructures.

\subsection{Parametric synchronization and parametric excitation in spin torque nano oscillators}
\label{Nov-Sync}

Besides the amplification of traveling waves, parallel parametric amplification has also been applied to spin-transfer torque driven nano-oscillators (STNO)\cite{StilesSTT, ZengSTNO,KiselevSTNO,RippardSTNO} to study the underlying physical phenomena and to improve their output characteristics. In an STNO, the magnetization of an active layer, typically referred to as the free layer, is subjected to an incident spin current. This spin current can be provided, for instance, by a direct current in a magnetoresistive device such as a spin valve or a magnetic tunneling junction\cite{KiselevSTNO,RippardSTNO,Dussaux-VSTNO,Durrenfeld20141,Quinsat-2012} or by the inverse spin Hall effect in an adjacent heavy-metal layer\cite{Demo-SHSTNO,Liu-SHSTNO}. Despite the various differences in geometry, magnetization configuration and material systems, all STNOs are based on the common principle that the incident spin current results in a spin transfer torque (STT)\cite{Slonczewski,Berger,StilesSTT} acting on the magnetization of the free layer which counteracts the damping. Similar to the impact of parametric amplification, this leads to a growth of the oscillation amplitude of the STNO if the inserted energy per unit time exceeds the loss rate of the STT-driven mode. Spin-wave amplification by STT is, thus, also a threshold process. This mechanism allows for the generation of an RF current by the conversion of a DC current. It should, however, be noted that it is hardly possible to select the fundamental mode to be the dominant mode amplified by STT in a spin-wave waveguide. Therefore, it is difficult to implement it as a means of spin-wave amplification in extended systems such as spin-wave waveguides\cite{Evelt-2016}. In contrast, it was demonstrated that it can constitute an efficient means to excite standing waves in a highly-quantized geometry in STNOs.

The manifold of device geometries for STNOs leads to a large span of frequencies which can be covered by such RF sources. This renders STNOs highly attractive for device applications, such as RF communications, where miniaturized RF sources with a large tunability are needed. However,  due to the lack of a frequency-selectivity, the DC STT interacts with the entire spin-wave spectrum. From a fundamental point of view this gives rise to many interesting nonlinear phenomena and creates a large playground to study non-linear spin-wave dynamics. For a device application, however, this hampers the functionality of the STNOs, since the interaction of the STT with the entire spin-wave spectrum leads to an undesired increase of the noise present in the oscillator. The resulting spin-wave--spin-wave interaction can result in a loss of phase-stability and increases the linewidth of the oscillator. Thus, typically STNOs are systems where the spin waves are strongly quantized and the spin-wave spectrum should be as simple and diluted as possible to ensure good device properties. Since this is not always easy to achieve, many STNOs suffer from a rather large linewidth. In addition, the output power of individual STNOs is too low for most practical applications.

One proposed way to overcome these issues is the synchronization of an array of oscillators. Such an array with oscillators working in parallel can produce a large output power where the mutual phase-locking of the oscillators provides the necessary phase-stability and reduces the linewidth\cite{Slavin2009}. First experimental evidence of phase-locked pairs of point contacts\cite{MancoffPL,KakaPL} has been followed by the demonstration of the synchronization of up to 5 oscillators\cite{Ruotolo2009,Sani2013,Houshang2016}. However, the creation of large synchronized arrays is a challenging task. This is, for instance, due to the limited reproducibility of the exact device shape, resulting in a frequency detuning between individual devices, and the large joule heating created in an array\cite{Slavin2009-2}. Moreover, due to the limited spin-wave propagation range, only oscillators situated close-by can be directly synchronized via spin-wave beams. In order to analyze the full potential and the limitations of phase-locking experimentally, as well as to explore the optimum operation conditions, large efforts have also been devoted to study the interaction of an STNO with an external locking-source, such as a microwave generator. Among these experiments, the parametric interaction of oscillators with a microwave current at twice their free-running auto-oscillation frequency has proven a vital technique, since it allows to decouple the driving frequency from the detected fundamental frequency of the oscillator. As will be briefly discussed in the following, these experiments have, among others, allowed to study the influence of noise and the applied pumping power on the locking bandwidth, have enabled an alternative quantification of the effect of the STT on the damping by analyzing its influence on the parametric instability threshold, and have proven that the coupling to an external pumping field can compensate the frequency mismatch of different oscillators in the locked state.

\begin{figure}[h]
\center
{\caption{First experimental observation of parametric generation in an STNO. a) BLS intensity as a function of the applied microwave power at the pumping frequency ($f$, black) and at the frequency of the parametrically generated spin waves ($f/2$, red). b) Influence of the applied DC bias on the threshold pumping power. c) Excitation spectrum for a numerical simulation of the magnetization dynamics if only the longitudinal ($h_x$, black curve) or transverse ($h_y$, red curve) component of the dynamic Oersted fields acts on the STNO. (Reprinted figures (a) and (b) from H. Schultheiss et al., Direct current control of three magnon scattering processes in spin-valve nanocontacts, Phys. Rev. Lett. 103, 157202 (2009). Copyright (2016) by the American Physical Society. Figure (c) reprinted from F. Ciubotaru et al., Mechanisms of nonlinear spin-wave emission from a microwave driven nanocontact, Phys. Rev. B 84, 144424 (2011). Copyright (2016) by the American Physical Society.)} 
 \label{Fig:Nov-PS1}}
{\includegraphics[width=1.0\textwidth]{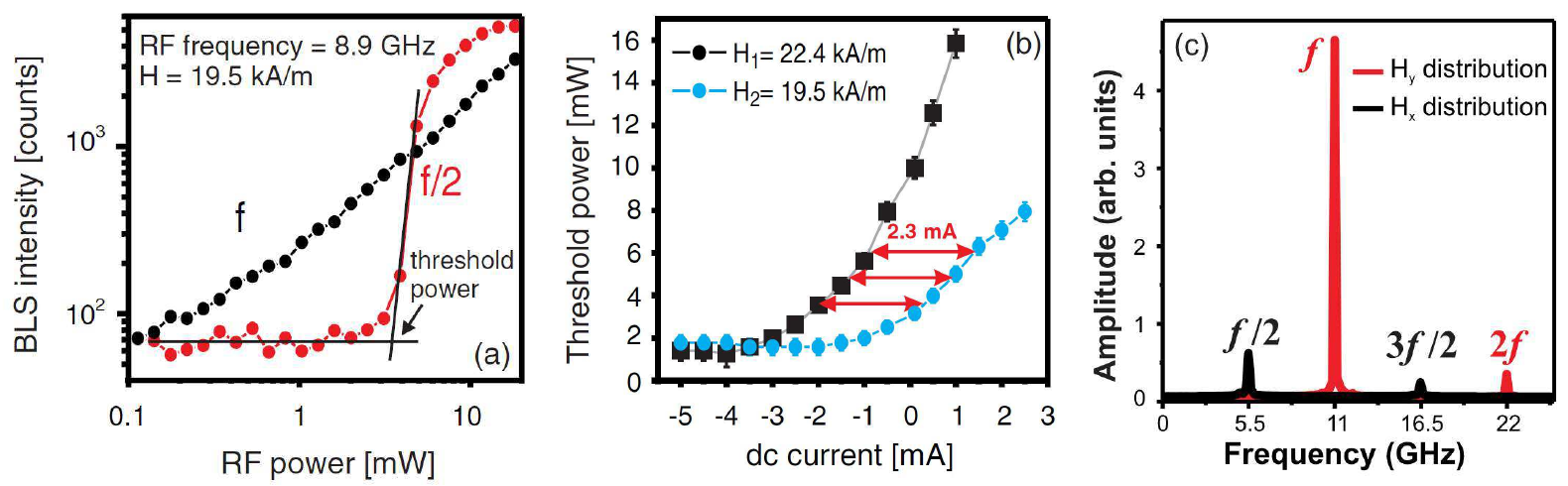}}
\end{figure}

The first observations of the excitation of spin waves by parallel pumping in a magnetic point contact has been made in Ref. \cite{Schultheiss-2009-1}, even though the proper identification of the parallel pumping process has been only made later in Ref. \cite{Florin-2011} by means of numerical simulations. In Ref. \cite{Schultheiss-2009-1}, the frequency-response of a Ni$_{80}$Fe$_{20}$ layer to a RF current which is injected by a point-contact of a diameter of about $200\,\mathrm{nm}$ has been studied by means of BLS microscopy. The Ni$_{80}$Fe$_{20}$ is the free layer in a Co$_{90}$Fe$_{10}$/Cu/Ni$_{80}$Fe$_{20}$ spin valve. The measurements have been performed at room temperature. It has been found that beyond a certain threshold value of the injected RF power a strong generation of spin waves at a frequency equal to one half of the frequency of the driving microwave current is observed. Figure \ref{Fig:Nov-PS1}~(a) shows the BLS intensity measured at the microwave frequency $f = 8.9\,\mathrm{GHz}$ and at one half of the externally applied frequency $f/2 = 4.45\,\mathrm{GHz}$ as a function of the applied microwave power in the absence of a DC bias. During the measurement, an external field of $\mu_0H_\mathrm{ext} = 24.8\,\mathrm{mT}$ is applied in-plane, perpendicular to the top electrode of the point contact ($x$-direction in Ref. \cite{Schultheiss-2009-1}). As can be seen from the figure, a clear threshold behavior is observed for the spin waves at $f/2$. Moreover, the intensity of the spin waves at frequency $f$ and, in particular, its increase with increasing microwave power, are not affected by the onset of the instability which leads to the formation of spin waves at $f/2$. This is strong evidence that the observed instability is indeed the parametric instability due to longitudinal parallel pumping: If the growth of the spin-wave intensity at $f/2$ would be related to a spin-wave instability - as has initially been assumed in Ref. \cite{Schultheiss-2009-1} - this should result in a kink in the dependence of the spin-wave intensity at $f$ as a function of the driving microwave power at the threshold of the instability\cite{SW_basics,Heusler-PRL}. After the onset of the three-magnon interaction, the intensity is supposed to grow with a smaller slope with respect to $f$. The absence of a three-magnon splitting in the experiment is further supported by the fact that this splitting process is impossible in the experimental conditions due to momentum conservation \cite{Florin-2011}. 

In Ref. \cite{Schultheiss-2009-1}, the influence of an additional DC current with current densities below the threshold of STT-driven auto-oscillations on the threshold of the parametric instability has been investigated. As can be seen from Fig.~\ref{Fig:Nov-PS1}~(b), a strong modulation of the threshold power as a function of the applied bias current is observed. The observed quadratic dependence of the threshold power on the applied DC current is consistent with the parametric instability: Since the STT is resulting in a linear change of the relaxation frequency\cite{Slonczewski,KiselevSTNO}, it leads to a linear change in the threshold field $h_\mathrm{th}$ (cf. Eq.~\ref{Eq:Threshold}). Since $P\propto h^2$, a quadratic change of the threshold power is expected. 

In Ref. \cite{Florin-2011}, the experimental configuration of Ref. \cite{Schultheiss-2009-1} has been investigated by means of micromagnetic simulations\cite{OOMMF}. In these simulations, the excitation mechanism has been indeed identified with the parametric instability and the Oersted field created by the current flow through the top contact was identified as the origin of the pumping field. This can be comprehended from Fig.~\ref{Fig:Nov-PS1}~(c), where the simulated excitation spectra are shown if the excitation is only created by the transverse Oersted field $h_y$, which can exert a direct torque on the magnetization, or by the longitudinal Oersted field $h_x$, which fulfills the requirements for parallel parametric amplification. The distribution of the Oersted field has been computed numerically. As can be seen from the figure, the transverse field only leads to an excitation at frequency $f$ and a higher harmonic at $2f$. In contrast, the longitudinal component, which cannot exert a direct torque on the magnetization, leads to the formation of spin waves at $f/2$. Since the spin waves at $f/2$ are observed exclusively in simulations where the longitudinal component is included and since no spin waves at frequency $f$ are excited under these conditions, a spin-wave instability can indeed be excluded as origin of the instability observed in the experiment. The fact that the excitation at $f$ and at $2f$ are mediated by the different components of the dynamic Oersted field also explains there complete independence in terms of their intensity-vs.-power characteristics (cf. Fig.~\ref{Fig:Nov-PS1}~(a)). 

Thus, the experiment in Ref. \cite{Schultheiss-2009-1} constitutes the first observation of parametric generation in a nanocontact and provides a first insight of the interplay of the damping change mediated by the DC-STT and the parallel parametric amplification in such a device. The parametric generation in a STNO below the threshold of the instability due to the STT has been labeled as \textit{parametric excitation} in Ref. \cite{Uraz-STNO-2010}. In this work, the parallel parametric amplification in a STNO was studied more systematically and the region below and beyond the threshold for the DC-STT driven auto-oscillations has been investigated experimentally as well as theoretically. The study was conducted on a Ni$_{80}$Fe$_{20}$/Cu/Co$_{70}$Fe$_{30}$ spin-valve, where the polarizing top Co$_{70}$Fe$_{30}$ and a part of the Cu layer were patterned into an ellipse of dimensions $100\times50\,\mathrm{nm}^2$. The microwave pumping field was provided by an additional micro-strip patterned on top of the nanocontact. In the study, an external bias field is applied in the sample plane at an angle of $45\,^\circ$ with respect to the long axis of the ellipse. The dynamic magnetization in the free layer is detected via the magnetoresistance. All measurements were performed at $5\,\mathrm{K}$. The applied microwave power is always below the parametric generation threshold in the absence of STT.

\begin{figure}[h]
\center
{\caption{a) Measured power spectral density for different applied pumping frequencies. The curves are shifted vertically for clarity. b) Full width have maximum of the excitation-peaks observed in (a) as a function of the pumping frequency. c) and d) Minimum and maximum locking frequency $f_\mathrm{min}$ and $f_\mathrm{max}$ as a function of the applied power (black circles: experimental values, red line: theoretical model). The circles/red curves include the locking range. In (c) the applied DC current of $I_\mathrm{DC} = 1.7\,\mathrm{mA}$ is below the auto-oscillation threshold, while in (d), the applied current of $I_\mathrm{DC} = 2.3\,\mathrm{mA}$ is larger than the threshold current of auto-oscillations. (Figures reprinted from S. Urazhdin et al., Parametric Excitation of a Magnetic Nanocontact by a Microwave Field, Phys. Rev. Lett. 105, 237204 (2010). Copyright (2016) by the American Physical Society.)} 
 \label{Fig:Nov-PS2}}
{\includegraphics[width=0.8\textwidth]{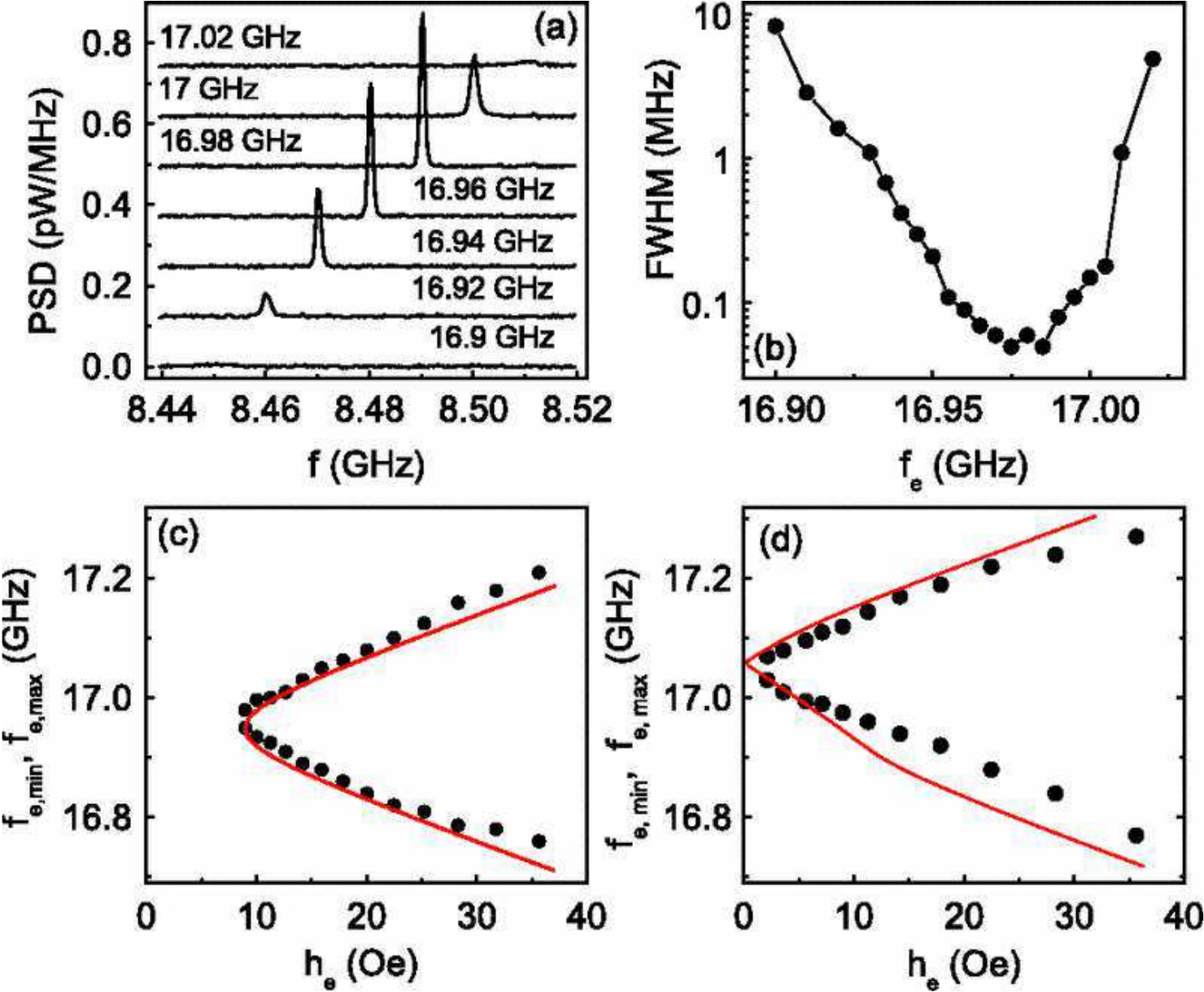}}
\end{figure}
Below the threshold of STT-driven auto-oscillations, parametric excitation can be used to study the spectral characteristics of the STNO and to study the damping change inflicted by the STT. This way, the parametric excitation of an STNO provides a powerful alternative to spin torque ferromagnetic resonance spectroscopy (ST-FMR)\cite{STFMR1,STFMR2}, since the microwave excitation at frequency $f$ and the dynamic response of the oscillator at $f/2$ are decoupled. Figure~\ref{Fig:Nov-PS2}~(a) shows the dynamic response of the STNO for different microwave frequencies for a fixed microwave field $\mu_0 h = 1.23\,\mathrm{mT} > h_\mathrm{th}(I_\mathrm{DC} = 1.7\,\mathrm{mA}) = 0.8\,\mathrm{mT}$ and DC bias $I_\mathrm{DC} = 1.7\,\mathrm{mA} < I_\mathrm{DC, th} = 2.0\,\mathrm{mA}$. As can be seen from the figure, a sizable parametric generation can be observed even for microwave frequencies different from the resonance frequency of the STNO of $f \neq f_0 = 8.465\,\mathrm{GHz}$ for the used magnetic bias field. This frequency detuning, which corresponds to the locking-range of a frequency-locked STNO according to Ref. \cite{Uraz-STNO-2010}, depends on the applied supercriticality of the pumping field\footnote{In principle, it also depends on the applied DC bias - the closer to the threshold of DC-STT driven auto-oscillations, the larger the locking-range at a given supercriticality in $h_\mathrm{RF}$.}. This can be comprehended from Fig.~\ref{Fig:Nov-PS2}~(c), where the locking-range is demonstrated as a function of the applied pumping field. The locking range is given by the area enclosed by the black circles, which correspond to the minimum and maximum frequency at which parametric excitation can still be observed experimentally, respectively. As can be comprehended from Fig.~\ref{Fig:Nov-PS2}~(a) and (b), the most efficient generation is achieved for $f/2 \approx f_0$, which is also associated with the minimum linewidth of the parametrically excited spin waves in the STNO. Due to the fact that in this scenario the STT has not compensated the spin-wave damping, a finite threshold field always remains for parametric excitation.

This is different in the case of parametric synchronization. If the DC bias is increased beyond the threshold of auto-oscillations, the 
oscillator can be locked without a threshold RF current (cf. Fig.~\ref{Fig:Nov-PS2}~(d), where the locking range is depicted as a function of the microwave pumping field above the DC STT-driven auto-oscillation threshold). Thus, beyond this threshold, the oscillator can easily be phase-locked to the parametric excitation source and a convenient \textit{parametric synchronization} to the external source can be observed, which has also been demonstrated by the same authors in Ref. \cite{Uraz-STNO-2010-1}. In the regime of parametric synchronization, the supercriticality of the pumping field mainly affects the locking-range of the oscillator.

These first experiments were followed by a multitude of experiments in different devices and geometries and the development of the necessary theoretical framework for their description. Besides studies on point-contacts\cite{Demi-2011-Control,Wang-2013,Rippard-2013}, where, for instance, the parametric generation and emission of propagating spin waves due to a frequency mismatch between the oscillator and the microwave pumping was demonstrated\cite{Demi-2011-Control}, experiments were extended to STNO incorporating magnetic tunneling junctions (MTJ)\cite{Quinsat-2011,Tiwari2016,Durrenfeld20141,Durrenfeld20142}, free layers in the magnetic vortex configuration\cite{Hamadeh-2014,Dussaux-2011,Bortoletti-2013,Martin-2011,Martin-2013,Lebrun-2015} and, more recently, to spin-Hall-effect driven STNOs\cite{Demi-STNOlock}. 

\begin{figure}[b!]
\center
{\caption{a) Top: Frequency of the double-vortex oscillator relative to one half of the driving frequency $f$. Bottom: Linewidth of the oscillator. The inset shows the device configuration schematically. The blue squares and dots represent experimental conditions with a low free-running linewidth of the oscillator and the red squares and dots with a large free-running linewidth (cf. inset in the top panel, where the free-running resonance peaks are shown). The bold solid lines in the top panel are theoretical predictions following Ref. \cite{Georges-2008}. b) Schematic of the parametrically synchronized spin-Hall-effect driven STNO studied in Ref. \cite{Demi-STNOlock}: Cr/Au leads are patterned as a bow-tie on top a Py/Pt disc. A DC current and an RF current are passed through this point-contact to excite magnetization dynamics in the STNO. c) Frequency of the oscillator as a function of the pumping frequency (blue squares), showing frequency locking in the shaded region. The white diamonds show the maximum output intensity (after \cite{Demi-STNOlock}). d) Intensity map of the nonlinear spin waves at $2f$ created within two constriction-based STNOs during auto-oscillation. The outline shows the extents of the STNOs. e) Frequencies of the oscillators as a function of the spacing between them (Figures (a) and (b) reprinted from A.~Hamadeh et al., Perfect and robust phase-locking of a spin transfer vortex nano-oscillator to an external microwave source, Appl. Phys. Lett. 104, 022408 (2014) with the permission of AIP publishing. Figures (c) and (d) reprinted by permission of Macmillan Publishing Ltd: [Nature Communications] (V. E. Demidov et al., Synchronization of spin Hall nano-oscillators to external microwave signals, Nat. Commun. 5, 3179 (2014)), copyright (2016). Figures (e) and (f) reprinted from T. Kendziorczyk and T. Kuhn, Mutual synchronization of nanoconstriction-based spin Hall nano-oscillators through evanescent and propagating spin waves, Phys. Rev. B 93, 134413 (2013). Copyright (2016) by the American Physical Society.)} 
 \label{Fig:Nov-PS3}}
{\includegraphics[width=1.0\textwidth]{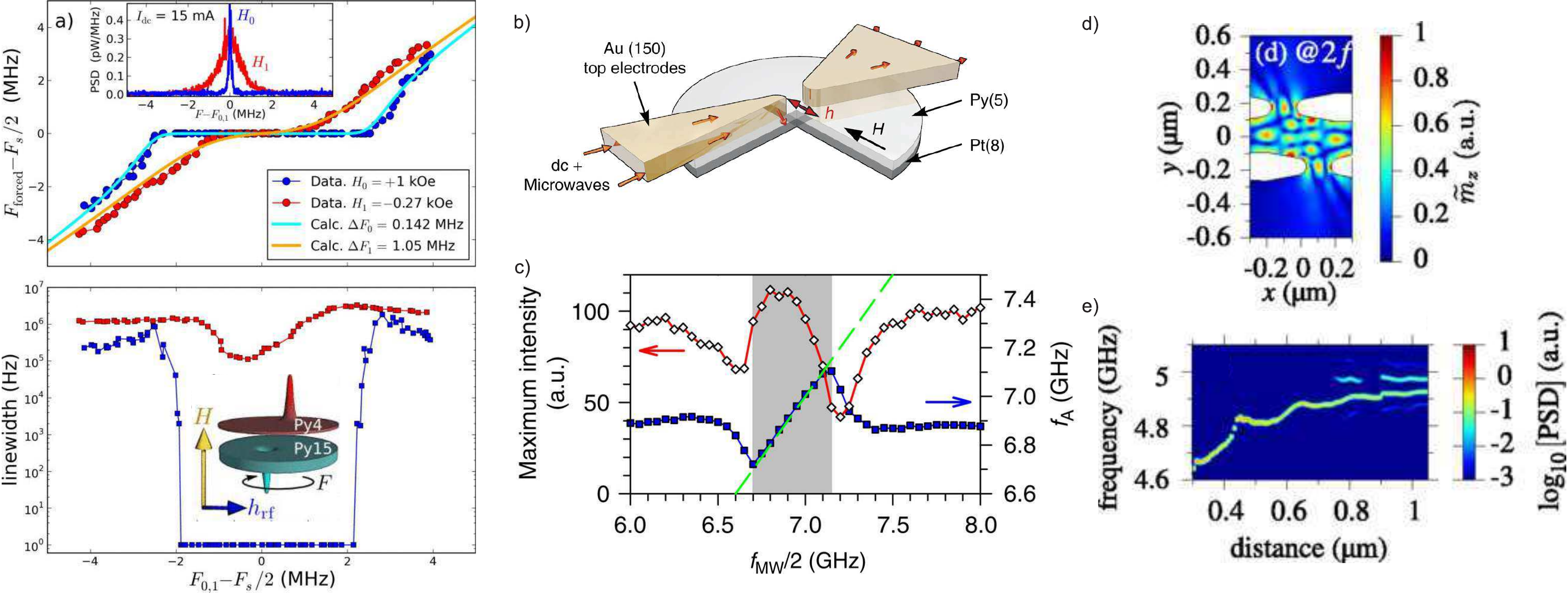}}
\end{figure}
For instance, in the early work on vortex-oscillators in Ref. \cite{Dussaux-2011}, it was demonstrated that parallel pumping can lock two oscillators which feature different free running auto-oscillation frequencies within the locking range. Moreover, in the more recent experiment in Ref. \cite{Hamadeh-2014}, the perfect phase-locking by parametric synchronization of a double-vortex oscillator, such as the one schematically shown as an inset  in the lower panel in Fig.~\ref{Fig:Nov-PS3}~(a), is demonstrated. The oscillator features a vortex in the free as well as the fixed layer, which mutually precess in the DC-STT driven auto-oscillating regime. Such devices can feature remarkably low linewidths in the free running regime at room temperature, such as the reported $\Delta f = 0.142\,\mathrm{MHz}$ relative to the free running frequency of $f_0 = 586\,\mathrm{MHz}$. The microwave pumping in this experiment is provided by a micro-strip antenna patterned on top of the STNO, which provides a microwave pumping field with frequency $f$ in the plane of the discs.

As demonstrated by the blue curve in Fig.~\ref{Fig:Nov-PS3}~(a), where the frequency of the STNO is shown relative to one half of the applied driving frequency $f$ of the pumping field, this enables the robust perfect phase-locking over a range $f_0-f/2$ over a locking-range greatly exceeding the free-running linewidth with a frequency linewidth $\Delta f_\mathrm{locked} < 1\,\mathrm{Hz}$. The importance of the phase-stability of the free-running oscillator is highlighted by the red curve: If the excitation field is changed, the mode-spectrum of the oscillator changes and the linewidth increases to a rather poor value of $\Delta f_0 = 1.05\,\mathrm{MHz}$ at a free running frequency of $684\,\mathrm{MHz}$. As is evident from the figure, this leads to a drastic decrease of robustness of the locking as well as of the best achievable linewidth, which increases to about $100\,\mathrm{kHz}$. This behavior was theoretically explained with the model presented in Ref. \cite{Georges-2008} (cf. bold lines in Fig.~\ref{Fig:Nov-PS3}~(a)).

More recent experiments on MTJ-based oscillators have demonstrated the robustness of parametrically synchronized STNO to a modulation of the applied microwave current if the modulation frequency is too fast for the nonlinearity of the oscillator to follow\cite{Durrenfeld20142}. Other experiments have demonstrated the possible use of a STNO to detect microwaves at the pumping frequency via spin rectification\cite{Tiwari2016}. All of these device geometries have in common that parametric synchronization does not show a threshold behavior and, in general, is connected to a significant increase of the peak spectral density (PSD) (i.e., of the intensity of the maximum of the peak). They also show that the linewidth of the peak under synchronization benefits from an initial low phase-noise of the free running oscillator (compare, e. g., Refs. \cite{Hamadeh-2014} and \cite{Quinsat-2011}). Further light was shed on this in the recent experiments on parametric synchronization of a spin-Hall-effect driven STNO\cite{Demi-STNOlock}. In this experiment, the microwave emission of such an STNO based on a Ni$_{80}$Fe$_{20}$/Pt bilayer, subjected to parametric excitation as well as parametric synchronization was characterized by means of BLS and by measuring the dynamic anomalous magnetoresistance (AMR)\cite{Liu-2013}. The device design is shown in Fig.~\ref{Fig:Nov-PS3}~(b). The RF field for the parallel pumping was provided by the part of the RF current sent through the oscillator flowing inside the Pt-layer. As shown in Fig.~\ref{Fig:Nov-PS3}~(c), where the frequency of the oscillator is shown as a function of the pumping frequency, it was observed that parametric synchronization of the STNO is possible. It was also found by the electrical measurements that it leads to a significant decrease of the linewidth of the oscillator. However, in this type of STNO, parametric synchronization requires a threshold microwave pumping power, has a comparably small locking-bandwidth and does not lead to a significant increase of the PSD. These findings have been explained in the framework of the theories provided in Refs. \cite{Slavin2009} and \cite{Georges-2008} and could be ascribed to the presence of a prominent thermal noise in the spin-Hall-effect driven STNO. It should be noted that the parametric excitation, or rather the effect of the STT on the parametric generation, has been studied by the same authors in Ref. \cite{Edwards-2012}. In this work, they demonstrate that the STT can be used to lower the generation threshold in a microscopic Ni$_{80}$Fe$_{20}$/Pt disc similar to the ones studied in Ref. \cite{Ulrichs} (cf. Section \ref{Pgen-Ell}) and that, in turn, parametric generation can be used to quantify the STT-induced damping change. This approach has also been successfully applied to study the influence of the spin-Hall effect driven STT in a macroscopic bilayer of YIG/Pt\cite{Vektor-2016}.

Another interesting prospect of parametric synchronization in spin-Hall-effect driven STNO was demonstrated in Ref. \cite{Kendziorczyk}. In this work, a single as well as a pair of nano-constriction based oscillators\cite{Demi-2014-STNO} were analyzed by micromagnetic simulations. A key to this numerical study is the lateral displacement of the two oscillators, which can be comprehended from Fig.~\ref{Fig:Nov-PS3}~(d), where a color map shows the spin-wave intensity at $2f$ and the outline shows the extents of the oscillators. These spin waves at $2f$ are created as a higher harmonic in the oscillator and lie within the band of propagating spin waves. They are created in the individual STNOs and are emitted at an oblique angle. If the design of the device is adapted and the spin waves emitted from one STNO can reach the second STNO and vice versa, an efficient mutual parametric synchronization of the STNOs mediated by these spin waves is possible. This is demonstrated in Fig.~\ref{Fig:Nov-PS3}~(e), where the spectrum of the oscillator pair is shown as a function of their distance. If they are far apart, the spin waves have decayed significantly during their propagation from one oscillator to the other. Consequently, their intensity is small to lock the oscillators and the individual spectra of the two STNO are visible. Since the geometric constrictions are slightly different, the  STNOs feature different resonance frequencies. Nevertheless, for a still substantial separation of about $800\,\mathrm{nm}$, an efficient spin-wave mediated locking of the oscillators is observed and they oscillate with a mutual frequency. For smaller distances, the locking becomes even more robust. The oscillatory behavior of the characteristics has been attributed to the interference of the short-wavelength spin waves at $2f$, which is visible in Fig.~\ref{Fig:Nov-PS3}~(d) and, at small separations, to the influence of the evanescent waves at frequency $f$ created in the STNOs. Hence, the parametric locking of the oscillators to the propagating spin waves at $2f$ allows for a mutual phase-locking of STNOs, even in an array with rather large device separations exceeding $500\,\mathrm{nm}$. 

\section{Concluding remarks}
\label{Concl}

In this Review we have given an overview over the recent advances in the application of the technique of parallel parametric amplification to magnetization dynamics in micro- and nanostructures. Parametric generation allows to study the dominant mode spectra and enables the efficient excitation of spin waves with very short wavelengths, practically independent of the size of the pumping source. This way, it gives access to plenty of nonlinear spin-wave phenomena and paves the way for the observation of Bose Einstein condensation of magnons in micro- and nanostructures made from materials with ultra low damping. Moreover, due to the fact that the frequency- and mode-selectivity of parametric amplification allow for the dominant amplification of the fundamental mode in these quantized systems, this method provides a very efficient means of spin-wave amplification. This new paradigm, which is very different from the previous applications of parallel pumping in macroscopic structures, allows for new applications of parallel pumping to create, amplify and manipulate information processed by magnons in microscopic devices. Its selectivity makes parallel pumping a superior method for the amplification of traveling waves, since the energy from the pumping field is not distributed over the entire spin-wave spectrum. However, there are still some issues to be addressed in this field: Most of the current demonstrations of parallel pumping have been based on the utilization of Oersted fields created by the RF currents sent through an adjacent micro-strip or through the structures itself. This approach is limited in scalability and requires quite large microwave fields. On the one hand, since the absorption of the individual structures is low and the applied microwave field can interact with a large number of structures, the field can be used as a global reference phase on the chip and it can be implemented as an active part in the logic circuits - benefiting in particular from the development of new materials with large saturation magnetization (i.e., large coupling) and low damping. Nevertheless, the interplay with individual nanostructures will require alternative approaches to create the pumping field in a local and energy-efficient way. Since any interaction which alters the effective field in parallel to the alignment of the static magnetization can be used, there are, however, plenty of options to achieve this. First hint that a part of the parametric excitation in STNOs are promoted at least partially by the AC-STT are just one instructive example of where the journey is heading. An identification of alternative driving mechanisms such as the variation of the anisotropy field by electric fields or mechanical stress in magnetostrictive layers, the utilization of field-like spin-orbit torques or of STT in the current-perpendicular-to-plane geometry will constitute the next milestones for the application of parallel parametric amplification. The feasibility of these means for the parametric spin-wave amplification needs to be explored experimentally as well as theoretically, and the corresponding coupling mechanisms need to be identified and optimized. In addition, the interplay of parametric amplification with a control of the magnetic damping by a DC-STT needs to be further explored and its full potential on the amplification of traveling waves needs to investigated. In the end, the journey will be a rewarding one: A decay-free propagation of spin waves in extended magnonic circuits becomes possible by parallel parametric amplification. The use of adiabatic and nonadiabatic amplifiers allows to satisfy all the needs for phase-dependent and phase-independent spin-wave amplification, adding vital ingredients to the magnonic 'toolbox'. 

\section*{Acknowledgments}
Financial support by the Graduate School Materials Science in Mainz (MAINZ) through DFG-funding of the Excellence Initiative (GSC 266) and by the DFG
(TRR49) as well as the Nachwuchsring of the TU Kaiserslautern is gratefully acknowledged.

\end{document}